\documentclass [amssymb,amsmath, twocolumn, showpacs]{revtex4-1} 
\usepackage{graphicx,epsfig,amsfonts,amssymb}
\usepackage{bm}
\usepackage{times}
\usepackage{lipsum}

\newcommand{\beq}{\begin{equation}}
\newcommand{\eeq}{\end{equation}}
\newcommand{\bes}{\begin{subequations}}
\newcommand{\ees}{\end{subequations}}
\newcommand{\bea}{\begin{eqnarray}}
\newcommand{\eea}{\end{eqnarray}}
\newcommand{\ba}{\begin{array}}
\newcommand{\ea}{\end{array}}
\newcommand{\beqn}{\begin{eqnarray*}}
\newcommand{\eeqn}{\end{eqnarray*}}

\newcommand{\ra}{\rangle}

\newcommand{\rar}{\rightarrow}

\begin{document}

\title{Exact results for models of multichannel quantum nonadiabatic transitions}

\author {N.~A. {Sinitsyn}$^{a}$ }
\address{$^a$ Theoretical Division, Los Alamos National Laboratory, Los Alamos, NM 87545,  USA}

\begin{abstract}
We consider nonadiabatic transitions in explicitly time-dependent systems with  Hamiltonians of the form $\hat{H}(t) = \hat{A} +\hat{B} t + \hat{C}/t$, where $t$ is time and $\hat{A}$, $\hat{B}$, $\hat{C}$ are Hermitian $N\times N$ matrices. 
We show that in any model of this type, scattering matrix elements satisfy nontrivial exact constraints that  follow from the absence of the Stokes phenomenon for solutions with specific conditions at $t \rightarrow -\infty$. This allows one to continue such solutions analytically to $t \rightarrow +\infty$, and connect their asymptotic behavior at $t \rightarrow -\infty$ and $t \rightarrow +\infty$. 
This property becomes particularly useful when a model shows additional discrete symmetries.  In particular, we derive a number of simple exact constraints and explicit expressions for scattering probabilities in  such systems.
\end{abstract}
\date{\today}

\maketitle

\section{Introduction}
 

 
Quantum nonadiabatic transitions have been studied for a long time with numerous applications in physics of atomic and molecular collisions \cite{book}. This field of research has strongly benefited from the discovery of exact  formulas that describe dynamics of two-state systems in specific but frequently encountered situations. The most famous such a theoretical result is the Stueckelberg-Majorana-Landau-Zener (LZ) formula \cite{maj,landau, LZ}. However, many other  exact results, such as the solution of the Rosen-Zener model and its generalizations \cite{rozen, nikitin, osherov}, have also been very influential and frequently used. 

More recently, the interest in quantum nonadiabatic transitions has been revived due to the new applications in  ultra-cold atomic systems \cite{app-bose,chain1}, quantum coherence \cite{coher}, Landau-Zener interferometry \cite{LZ-interferometry}, and quantum control of mesoscopic systems \cite{qcontrol}, which 
typically deal with quantum systems of mesoscopic size and a large phase space.  

The multistate version of the LZ model  is one of the most frequently emerging problems in these studies \cite{book}. 
It considers interactions among $N$  states during the time evolution described by the Sch\"odinger equation with  time-dependent parameters that change according to simple power laws. Specifically, here we will discuss the evolution equations of the form:
\begin{equation}
i\frac{d\psi}{d t} = \left( \hat{A} +\hat{B}t +\frac{ \hat{C}}{t} \right)\psi,
\label{mlz}
\end{equation} 
where $\psi$ is the state vector in a space of $N$ states; $\hat{A}$, $\hat{B}$ and $\hat{C}$ are constant Hermitian $N\times N$ matrices.


In this article, we will assume that matrices are written in the, so-called, {\it diabatic basis}, in which the matrix $\hat{B}$ is diagonal and if some of its diagonal elements $\beta_i$, $i=1,\ldots, N$, are degenerate then diabatic basis states are chosen to make constant couplings among such states equal to zero, i.e. in the diabatic basis we have: 
\beq
\hat{B}={\rm diag}\{\beta_1,\ldots \beta_N\}, \quad A_{nm}=0 \,\,\, {\rm if} \,\,\, \beta_n=\beta_m.
\label{diab}
\eeq
 Diagonal elements of the Hamiltonian
\beq
\hat{H}(t) = \hat{A} +\hat{B} t + \hat{C}/t
\label{masterH}
\eeq
 can be generally written  in the diabatic basis as
\beq
\varepsilon_i^ d= \beta_i t +\epsilon_i +k_i/t, \quad i=1,\ldots N,
\label{diab1}
\eeq
 where  $k_i$ are diagonal elements of $\hat{C}$ and $\epsilon_i$ are diagonal elements of $\hat{A}$. Off-diagonal elements of $\hat{A}$ and $\hat{C}$ in the diabatic basis are called the {\it coupling constants}.

 The goal of the theory is to find the scattering $N\times N$ matrix $\hat{S}$, whose element $S_{nn'}$ is the amplitude of the diabatic state $n$ at $t  \rightarrow +\infty$, given that at $t \rightarrow 0_{+}$ the system was in the $n'$-th eigenstate  of the Hamiltonian. 
In most cases, only the related matrix $\hat{P}$, $P_{n' \rightarrow n}=|S_{nn'}|^2$,  called the matrix of {\it transition probabilities}, is of interest.

Even the case of  Eq.~(\ref{mlz}) for only two states ($N=2$)    generally does not reduce to the hypergeometric equation, and its analytical solution, e.g. in the form of a contour integral of a simple function, is unknown. 
Situation may look even less promising at larger $N$ because Eq.~(\ref{mlz}) is then equivalent to an $N$-th order differential equation with  polynomial time-dependent  coefficients that quickly grow in complexity.

Although the general solution of the model (\ref{mlz}) has not been found, a number of exactly solvable cases with specific forms of matrices $\hat{A}$, $\hat{B}$ and $\hat{C}$ have been derived. Exact results provided useful intuition about the behavior of  strongly driven quantum systems. The models of type (\ref{mlz}) with $\hat{C}=0$  have been discussed rather extensively  in the past \cite{be,shytov,no-go,mlz-1,do, reducible,bow-tie,chain}.
In contrast, the addition of the last term in (\ref{mlz}), with $\hat{C}\ne 0$, has been introduced relatively recently in physics literature \cite{coulomb1}, and it will be the main focus of the present work.
  
We will refer to an arbitrary model of the type (\ref{mlz}) with a nonempty matrix $\hat{C}$ as an {\it LZC-model}, named after Landau, Zener and Coulomb. Exactly solvable LZC models have been shown to capture  very complex patterns of behavior, including counterintuitive transitions and wild oscillations of transition probabilities as functions of parameters \cite{sinitsyn-13prl,sinitsyn-13jpa,sinitsyn-14jpa}. Physically, they show a lot of common features with nonadiabatic transitions in Rydberg atoms \cite{stark} and molecular collision models \cite{singular}.



{\bf The goal of this article} is to explore an unusual phenomenon that appears to be common for all models of the type (\ref{mlz}). We will demonstrate that no matter how big is the complexity of such a model, there are always nontrivial but simple exact constraints on its scattering matrix elements in addition to the trivial constraints that follow from the unitarity and elementary  discrete symmetries. We will provide  tests of such constraints by numerical simulations and comparisons with available exact results, and demonstrate how they can be used to derive new relations between transition probabilities in specific systems.

The structure of our article is as follows. In section 2, we will explore the Stokes phenomenon in systems of the type (\ref{mlz}) and derive the exact constraints for the 
scattering matrix for such systems.  In section 3, we explore the case $N=2$ in more detail and demonstrate how constraints on the scattering matrix can lead to constraints on transition probabilities in elementary LZC-type models. Section 4 describes
 specific examples with $N>2$ for which simple constraints on transition probabilities can be derived. We summarize our results in the conclusion section 5, in which we also discuss possible directions for the future research.  Appendix A connects our results with previous studies of the special case of $\hat{C}=0$ in (\ref{mlz}). Appendix B presents the derivation of exact transition probabilities in a specific model with arbitrary $N$ that we use to check the validity of some of our results in the main text.

\section{Asymptotic behavior of extremal amplitudes in LZC models}

Our treatment of the  general case of the LZC model will closely follow the proof of the Brundobler-Elser formula and derivation of the no-go theorem in \cite{no-go}. The major difference from that work is that now we include  a nonzero value of the matrix $\hat{C}$ into account.  
Let us define the {\it extremal amplitude} as the amplitude of the diabatic state that has the highest  or the lowest slope at $t \rar \pm \infty$ i.e. that has the largest or the lowest eigenvalue of the matrix $\hat{B}$. 
\begin{figure}
\scalebox{0.175}[0.175]{\includegraphics{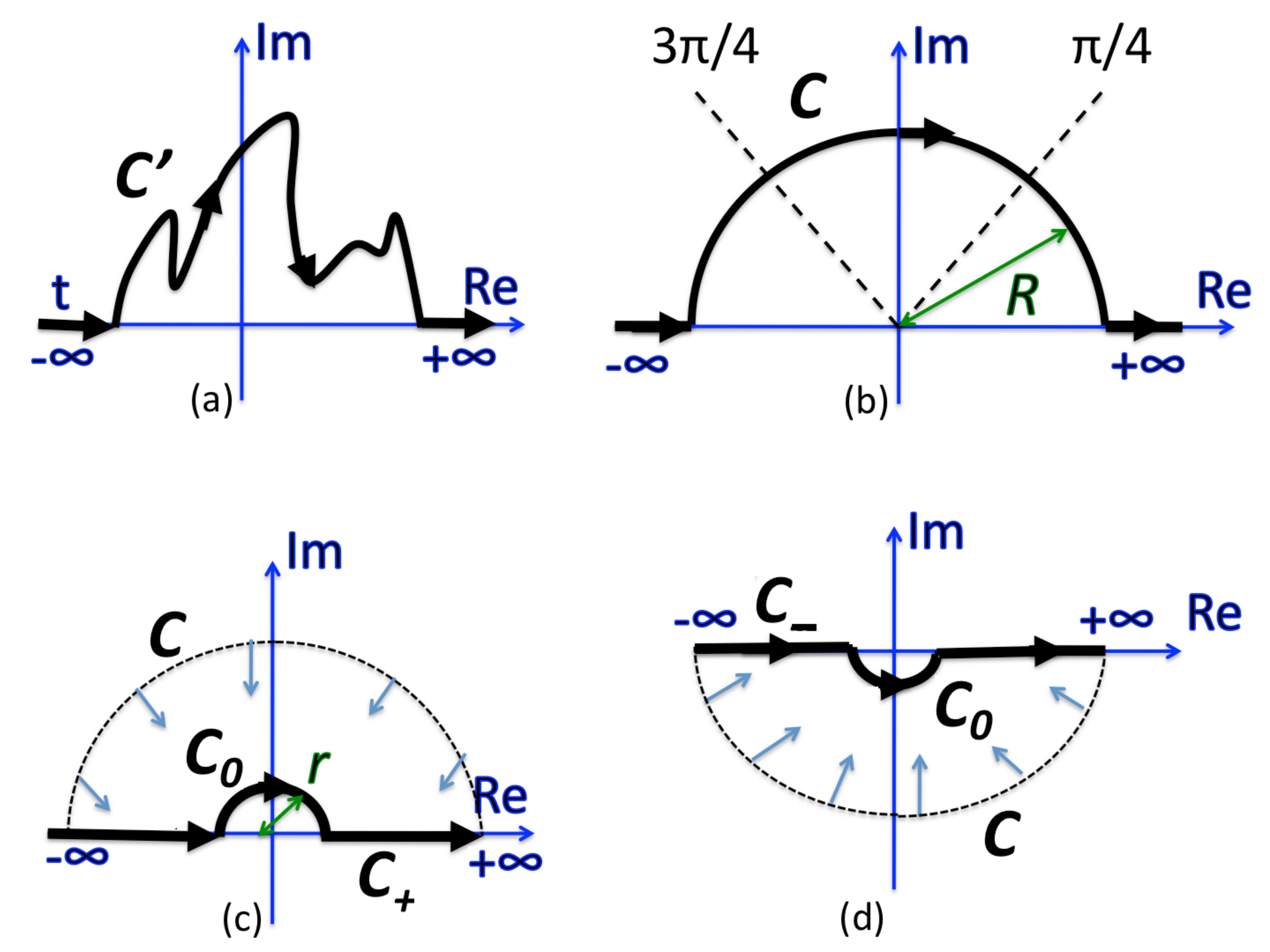}}
\hspace{-2mm}\vspace{-4mm}   
\caption{Contours of complex time for evolution with Eq.~(\ref{mlz}) connecting real points at $t\rightarrow \pm \infty$. (a) Arbitrary contour ${\bm C'}$ laying in the upper half plane can be deformed without encountering singularities into the contour ${\bm C}$ shown in (b) that has the shape of a semicircle with radius $R$. This contour crosses two rays, $t=Re^{i\phi}$, at $\phi=3\pi/4$ and $\phi=\pi/4$. Along those rays,  the highest slope amplitude has the highest rate of, respectively, decay and growth with increasing $R$. (c) Without changing asymptotic values of the solution at real $t\rar \pm \infty$, the contour ${\bm C}$ can be deformed to the contour that is placed almost everywhere on the real time axis, except an infinitely small semicircle  that avoids the singularity of the Hamiltonian (\ref{masterH}) at $t=0$. (d) Applying the same arguments to the extremal amplitude with the lowest slope of diabatic energy, we arrive at a contour ${\bm C}$ at the lower complex half-plane that can be deformed to a contour along the real axis except an infinitely small semicircle around $t=0$ in the lower half-plane.  }
\label{crossf}
\end{figure}
We can now prove the following rule:

\subsection{Connection formula}

Without loss of generality, let $1$ be the index of the extremal slope with $\beta_1=\max(\beta_{1} \ldots \beta_{N})$ and let 
${\bm C'}$ be an arbitrary contour that connects asymptotic values $t\rightarrow \pm \infty$ on the real axis but otherwise it makes arbitrary continuous path in the upper half of the complex plain avoiding the singular 
point of the Hamiltonian at $t=0$, 
  as shown in Fig.~\ref{crossf}(a). Suppose also that at real $t\rightarrow -\infty$ the asymptotic values of the amplitudes of diabatic states are given by
\beq
|\psi_1^{-\infty}(t)| = 1,\quad |\psi_{i}^{-\infty}(t)| = 0, \quad i\ne 1.
\label{bound1}
\eeq
We are going to show now that  the value of the extremal amplitude at real $t \rightarrow +\infty$ is given by
\begin{eqnarray}
 \psi_1^{ +\infty} (t)&=&S^{\rm up}_{11} \psi_1^{-\infty}(t) , \\
 S^{\rm up}_{11}&=& \exp \left(-\pi k_1- \pi \sum \limits_{i\,(i \ne 1)} \frac{|A_{1i}|^2}{|\beta_1-\beta_i|}  \right),
\label{connect1}
\end{eqnarray}
where parameters are introduced in Eqs.~(\ref{mlz})-(\ref{diab1}), and index ``up" indicates that evolution went along the time contour in the upper half of the complex plane.

Respectively, let  $N$ be the index of the extremal slope with $\beta_{N}=\min(\beta_{1} \ldots \beta_{N})$ and ${\bm C'}$ connects real $t \rightarrow \pm \infty$ in the lower half of the complex plane with initial conditions 
\beq
|\psi_N^{-\infty}(t)| = 1,\quad |\psi_{i}^{-\infty}(t)| = 0, \quad i\ne N,
\label{bound2}
\eeq
then 
\begin{eqnarray}
\psi_N^{ +\infty} (t)&=&S^{\rm dn}_{NN} \psi_N^{-\infty}(t),\\
 S^{\rm dn}_{NN} &=& \exp \left(+\pi k_N- \pi \sum \limits_{i\,(i \ne N)} \frac{|A_{Ni}|^2}{|\beta_i-\beta_N|}  \right),
\label{connect2}
\end{eqnarray}
where the index ``dn" indicates that evolution went along the time contour in the lower half of the complex plane.

\underline{\it Proof:} Consider  nodegerate $\beta_i$, $i=1,\ldots N$. We will prove the case of the highest slope first. 
 Since we are interested in the asymptotic
  magnitude of the amplitudes of states at large time, we do not have to find the evolution matrix for Eq.~(\ref{mlz}) at arbitrary time point of ${\bm C'}$. Moreover, since  the solution is analytic everywhere except $t=0$, 
  the deformations of the contour that do not change its asymptotic poins at the real axis do not change asymptotics of the solution. 
   Hence, we can analytically extend the evolution (\ref{mlz}) to  the path ${\bm C}$ that 
  always has $|t|\rightarrow \infty$, as shown in Fig.~\ref{crossf}(b) i.e.
  \beq
  {\bm C}: t=Re^{i\phi}\quad R\rightarrow \infty, \quad \phi \in [\pi, 0], 
  \label{cont1}
  \eeq
  where $R$ is real and positive.

\begin{figure}
\scalebox{0.185}[0.185]{\includegraphics{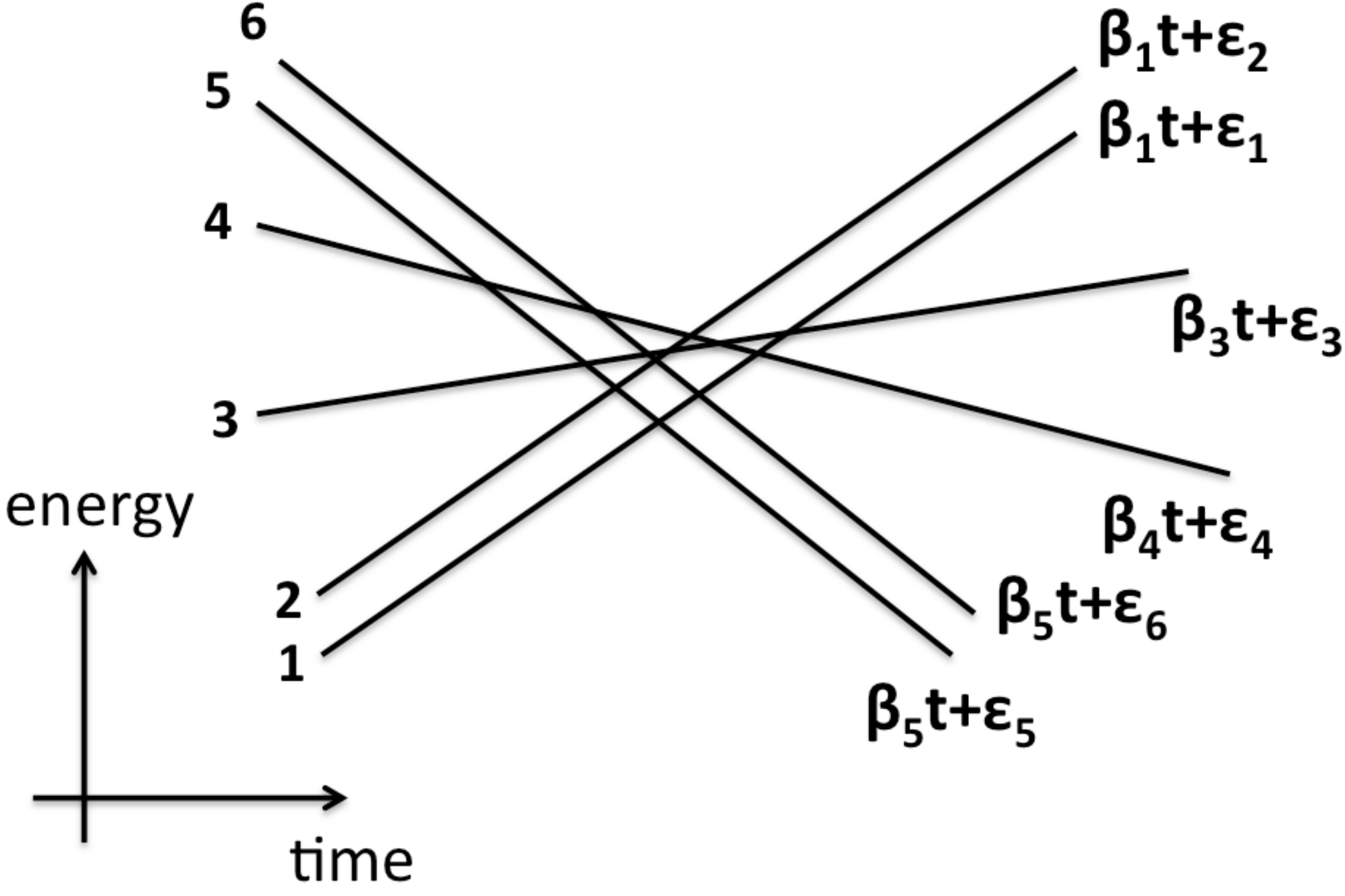}}
\hspace{-2mm}\vspace{-4mm}   
\caption{Plot of diagonal elements of the matrix $\hat{B}t+\hat{A}$ in the diabatic basis can be used to identify extremal states and counterintuitive transitions. For the shown system of six states, levels 1 and 2 have the highest slope, levels 5 and 6 have the lowest slope, and transitions from level 6 to level 5 and from level 1 to level 2 are counterintuitive.}
\label{levels1}
\end{figure}

 Along  the contour ${\bm C}$,  diabatic states coincide with eigenstates of the Hamiltonian. The distances between corresponding
  instantaneous eigenenergies $E_i(t)$ of the Hamiltonian (\ref{masterH}) remain always large in this case,
  namely of the order
  of $|(\beta_i-\beta_j)t|>>|A_{ij}|,\, |C_{ij}/t|$ for the states $i \ne j$ and hence one can use the adiabatic approximation for the amplitudes $\psi_i$ of the diabatic states: 
  \begin{equation}
  \psi_i(t)=e^{-i\int_{-\infty  }^{t} E_i(t)dt}\psi_i(-\infty ), \quad i=1,\ldots, N.
  \label{adiab}
  \end{equation}
To  the leading orders in $1/R$, the Hamiltonian eigenenergy that corresponds to the amplitude of the highest slope is given by
\beq
E_1 = \beta_1 t +\frac{1}{t} \left( k_1+ \sum \limits_{j\,(j \ne 1)} \frac{|A_{1j}|^2}{|\beta_1-\beta_j|} \right) + O(1/R^2).
\label{en1}
\eeq 
Now, as it was done in \cite{shytov,no-go} for the case of $\hat{C}=0$, we make the observation that by substituting (\ref{en1}) into (\ref{adiab}) and integrating over time along the semi-cycle $t=Re^{i\phi}$, where $\phi \in [\pi,0]$, we obtain (\ref{connect1}).

This observation, however, cannot be considered as the proof yet because, generally, the adiabatic approximation (\ref{adiab}) may break down for complex values of time, which is the essence of the Stokes phenomenon. 
The evolution along a  complex time path is no longer unitary so that some of the amplitudes can become exponentially large in comparison to $\psi_k$. In such a case, approximation (\ref{adiab}) cannot be applied because even  
a weak coupling to states with exponentially large amplitudes cannot be treated perturbatively. 

 In principle, to justify (\ref{adiab}), one can apply arguments akin to Landau's derivation of the LZ-formula and the treatment of over-barrier transitions \cite{landau}, as it was suggested in \cite{shytov} for systems with $\hat{C}=0$. However, such arguments are intrinsically semiclassical and generally predict only the leading exponential factor for a transition amplitude. For example, the semiclassical formula for the over-barrier reflection fails in the limit of a weak barrier, i.e. in the domain of applicability of the Born approximation. In order to make exact statements, one has to explore the Stokes phenomenon in this problem in more detail.

In order to prove that the perturbative expansion (\ref{en1}) can be used in Eq.~(\ref{adiab}) everywhere along the path ${\bm C}$, we should show that the amplitude with energy $E_1$ remains either exponentially larger or at least of the same order with other state amplitudes. The latter means here that the ratio of the extremal amplitude to any other one is not suppressed exponentially in the limit $R\rightarrow \infty$.  

It is sufficient for this proof to consider only  the leading order terms  in the exponents
 \beq
 \psi_i \sim \exp(-i\beta_it^2/2), \quad i=1,\ldots, N.
  \label{as1}
  \eeq
 Stokes phenomenon, i.e. sharp changes of the behavior of the values of the amplitudes at $R\rightarrow \infty$, can happen at crossing the Stokes lines, i.e. rays along which  some of the amplitudes are growing/decaying with extremal rate. In our case, those lines are the  rays  $\phi=\pi/4$ and $\phi=3\pi/4$. It is sufficient to prove that the extremal amplitude behaves continuously at crossing those rays.

 Consider the ray at $\phi=3 \pi /4$ and assume that  $\psi_1$ is of the same order or exponentially larger than all other state amplitudes at some very large $R$. This is the ray of the slowest growth of the extremal amplitude. 
Hence, by continuing asymptotics (\ref{as1}) to larger values of $\phi$, the amplitude $\psi_1$ is exponentially growing in comparison to 
all other amplitudes up to $\phi=\pi$. This means that this solution can be normalized to satisfy initial conditions (\ref{bound1}) and the extremal amplitude does not become suppressed in comparison to other states in the sector ($\pi, 3\pi/4$). Similarly, moving to the right from the ray $\phi=3 \pi /4$, we find that up to the ray $\phi=\pi/4$, the amplitude $\psi_1$ is exponentially growing in comparison to other amplitudes.  Finally, at interval $\phi \in (\pi/4, 0)$ other amplitudes start growing but from exponentially suppressed value at $\phi=\pi/4$. Therefore, it is safe to continue the  extremal amplitude analytically to the right of this ray and hence to the whole sector up to $\phi=0$. This completes our proof of the absence of the Stokes phenomenon for the extremal amplitude along the contour ${\bm C}$ and initial conditions (\ref{bound1}).  The proof for the level with the lowest slope is analogous but with the contour ${\bm C}$ placed in the lower half of the complex plane.

\subsection{No-Go Rule for LZC models}

Equations (\ref{connect1}) and (\ref{connect2}) are generalizations of the Brundobler-Elser formula in multi-state Landau-Zener models with linear time dependence of parameters, which we review briefly in Appendix A.
Our inclusion of the Coulomb term in (\ref{mlz}) changed the result but did not change basic steps discussed in \cite{no-go} for derivation of the connection formula for the case $\hat{C}=0$. Indeed, at large time values, the Coulomb term introduces only marginally relevant contribution, which produces a geometric-phase-like pre-factor,  which does not influence the  Stokes phenomenon. This observation can be used to derive another exact constraint. Namely, in  \cite{no-go},  the so-called ``No-Go Rule" was derived for the case when instead of one state with the highest (or one lowest) slope of the diabatic energy level there is a band of an arbitrary number of states having the same 
  highest slope  so that diabatic energies in this band are different only by constant energy parameters, as shown in Fig.~\ref{levels1}.
  
  \underline{\it The No-Go Rule} states that the so-called {\it counterintuitive} transitions are exactly forbidden. 
  Generally for the model (\ref{mlz}), if $\beta_m=\beta_n=\max(\beta_{1} \ldots \beta_{N})$
   then the transition from the state  $m$ to the state 
  of the same band $n$ is defined to be ``counterintuitive" if  $\epsilon_m<\epsilon_n$, where $\epsilon_i$ is defined in (\ref{diab1}).
   Correspondingly, if 
  $\beta_n=\beta_m=\min(\beta_{1} \ldots \beta_{N})$ then the transition is counterintuitive if  $\epsilon_{m}>\epsilon_{n}$. 
  For example, in Fig.~\ref{levels1}, transitions from the state $1$ to the state $2$ and from the state $6$ to the state
  $5$ are counterintuitive. Note that this definition allows arbitrary form of the matrix $\hat{C}$. 
  
  According to the No-Go Rule, the amplitude of a counterintuitive transition is vanishingly small, i.e.  for a specific element of the evolution matrix, we have
     \begin{equation}
S^{\rm up/dn}_{nm} =0
  \label{eq3}
  \end{equation}
when $n$ and $m$ are extremal amplitudes, and transition from $m$ to $n$ is counterintuitive in the sense described above, and the choice of ``up" or ``dn" index is 
according to whether levels $m$ and $n$ have, respectively, the highest or the lowest slope.

The no-go rule (\ref{eq3}) was proved in \cite{no-go} by exploring the Stokes phenomenon and therefore it is equally valid for the LZC model (\ref{mlz}). 
Indeed, suppose that we have more than one states with the highest slopes. Then the magnitudes of their amplitudes at the contour ${\bm C}$ are dominated by the exponents:
\beq
\psi_{n,m} \sim  \exp(-i\epsilon_{n,m} t-i\beta_1 t^2/2).
\label{expa}
\eeq

 Along the ray $\phi=\pi/2$,  the amplitude $\psi_n$ is growing faster than $\psi_m$ when $R$ is growing because $\epsilon_n>\epsilon_m$.
Consider a situation when for some very large $R$ we have the boundary condition that both amplitudes are comparable along this ray. Then the solution with $\epsilon_m<\epsilon_n$ is growing if $\phi$ is continued either to the right or to the left along the contour ${\bm C}$. Hence, it is possible to normalize this solution so that at real $t \rightarrow -\infty$ the state $m$ has a unit amplitude and state $n$ is vanishing. Continuation to the real positive time, $t \rightarrow +\infty$, will produce that $\psi_m$ is not influenced by state $n$ and satisfies (\ref{connect1}) and $\psi_n$ has a vanishing amplitude.

\subsection{Constraints on quantum mechanical evolution operator}

The connection formulas (\ref{connect1})-(\ref{connect2}), as well as the no-go rule (\ref{eq3}) apply  to arbitrary contour that can be obtained by a continuous deformation of a contour ${\bm C}$  without crossing the singular point at $t=0$. Therefore, at least part of this contour has to lie at nonzero imaginary part of the time $t$, which makes the evolution along this contour non-unitary. 
However, depending on whether ${\bm C}$ is in the upper or the lower parts of the complex plain,  one can deform ${\bm C}$ into one of the contours, either ${\bm C_+}$ or ${\bm C_{-}}$,  as shown in Fig.~\ref{crossf}(c,d), such that during real time intervals 
$t \in (-\infty,-r)$ and $t \in (r,+\infty)$   the evolution is unitary. We will choose to connect those intervals by a half-circle path ${\bm C_0}$ and assume that the radius $r$ of this path is infinitesimally small

  \beq
  {\bm C_0}: t=r e^{ \pm i \phi}, \quad r\rightarrow 0, \quad \phi \in [\pi, 0], 
  \label{cont1}
  \eeq
  where $r$ is real and positive and the choice of the sign of the phase corresponds to the choice of the semi-plane of the contour ${\bm C}$.
  
  In the limit $r \rightarrow 0$, the evolution along ${\bm C_0}$ is totally dominated by the singular term with matrix $\hat{C}$ in the Hamiltonian (\ref{masterH}). Hence the evolution operator over ${\bm C_0}$ can be easily found:
   \beq
 \hat{S}_0^{\rm up/dn} = \exp \left( {-i\int_{\bm C_0} dt \frac{ \hat{C}}{t} }\right)=\exp \left( \mp \pi \hat{C} \right),
 \label{s0}
 \eeq
 where ``up" and ``dn" indexes refer to the contour ${\bm C_0}$ is placed, respectively,  above or below the real axis. 
  
Let  $\hat{S}^{\rm up}$ and $\hat{S}^{\rm dn}$ be the $N\times N$ matrix scattering operators for evolution along the contours, respectively, ${\bm C_{+}}$ and ${\bm C_{-}}$,
illustrated in Fig.~\ref{crossf}(c,d); and let $\hat{S}_{-}$ and $\hat{S}_{+}$ be the operators for 
unitary quantum mechanical evolution  along the 
real time during intervals, respectively, $t \in (-\infty, 0_{-})$ and $t\in (0_{+},+\infty)$. Then:

\beq
\hat{S}^{\rm up/dn} = \hat{S}_{+}  \hat{S}_0^{\rm up/dn}  \hat{S}_{-},
\label{evol1}
\eeq
so that connection formulas and the no-go rule can be expressed as follows:
\beq
\left[ \hat{S}_{+}  \hat{S}_0^{\rm up}  \hat{S}_{-} \right]_{11} = \exp \left( -\pi k_1- \pi \sum \limits_{i\,(i \ne 1)} \frac{|A_{1i}|^2}{|\beta_1-\beta_i|}  \right),
\label{connect4}
\eeq
\beq
\left[ \hat{S}_{+}  \hat{S}_0^{\rm dn}  \hat{S}_{-} \right]_{NN} = \exp \left( \pi k_1- \pi \sum \limits_{i\,(i \ne N)} \frac{|A_{Ni}|^2}{|\beta_N-\beta_i|}  \right), \label{connect41}
\eeq  
 \beq 
 \left[ \hat{S}_{+}  \hat{S}_0^{\rm up/dn}  \hat{S}_{-} \right]_{nm} =0,
\label{connect42}
\eeq
where the transition from state $m$ to state $n$ is counterintuitive in the sense that was defined in previous subsection.

Equations~(\ref{s0})-(\ref{connect42})  are the  most general central result of this work. They say that for any Hamiltonian (\ref{masterH}) there are exact nonperturbative constraints on the scattering matrices $\hat{S}_+$ and $\hat{S}_{-}$ of the quantum mechanical evolution. At current stage, those constraints are not looking particularly useful because they do not provide an explicit expression for any particular matrix element of the physically useful scattering matrices 
 $\hat{S}_{\pm}$. In fact, constraints (\ref{connect4})-(\ref{connect42}) are expressed via the products of scattering matrices that describe evolution over disjoined time intervals.
 
 Nevertheless, we will show that  Eqs.~(\ref{s0})-(\ref{connect42})  become quite useful when sub-classes of LZC models with specific discrete symmetries are considered. In those cases, it is possible to connect the elements of 
  $\hat{S}_{-}$ with elements of  $\hat{S}_{+}$, and hence rewrite the matrices $\hat{S}^{\rm up/dn}$ only in terms of $\hat{S}_{+}$. After this, Eqs.~(\ref{connect4})-(\ref{connect42}) usually can be expressed as nontrivial constraints on the desired transition probabilities   between states of an LZC system during the evolution in $t \in (0_{+}, +\infty)$.
  
  

\section{Two-state systems}


The goal of this section is to provide elementary demonstrations of how transition probabilities in LZC models can be found  by using connection rules (\ref{connect4})-(\ref{connect42}).



\subsection{Case 1: Diagonal $\hat{C}$ and off-diagonal $\hat{A}$}
Consider the following evolution of two states:

\beq
i\frac{d}{dt} \left( \ba{c}
a\\
b
\ea \right)
= \left( \ba{cc}
k/t & g\\
g & \beta t
\ea \right) \left( \ba{c}
a\\
b
\ea \right).
\label{hh1}
\eeq  
Equation~(\ref{hh1}) is symmetric under  simultaneous

(i) reflection of time: $ t\rightarrow - t$, and

(ii) change of  the sign of the first amplitude: $a(t) \rightarrow -a(t)$.  

In terms of the evolution matrices, symmetries (i)-(ii) mean that if we write the evolution operator from $t=0_+$ to $t=+\infty$ in the matrix form,
\beq
 \hat{S}_{+}  \equiv \hat{S}(+\infty| 0_+) =  \left( \ba{cc}
s_{11} & s_{12}\\
s_{21} & s_{22}
\ea \right),
\label{sp1}
\eeq
then the evolution operator for backward in time evolution, starting from $t=0_{-}$ and ending at  $t=-\infty$ is given by 
\beq
\hat{S}(-\infty| 0_- ) =  \left( \ba{cc}
s_{11} & -s_{12}\\
-s_{21} & s_{22}
\ea \right).
\label{sm1}
\eeq

(iii) Finally, we recall the  symmetry, which is always present. Due to the unitarity, backward and forward in time evolutions are related by  complex conjugation of the evolution matrices, i.e. 
\beq
\hat{S}(-\infty| 0_- ) = \hat{S}^{\dagger} ( 0_-| -\infty ) .
\label{sm2}
\eeq

Since  $\hat{S} ( 0_-| -\infty ) \equiv \hat{S}_{-}$, (i)-(iii) mean that 
\beq
 \hat{S}_{-} =  \left( \ba{cc}
s_{11}^* & -s_{21}^*\\
-s_{12}^* & s_{22}^*
\ea \right).
\label{sp1}
\eeq

Consider the case with $\beta>0$. 
The evolution over the infinitesimal contour  around $t=0$  below the real axis gives:
\beq
 \hat{S}_0^{\rm dn} =   \left( \ba{cc}
e^{\pi k} & 0 \\
0 & 1
\ea \right).
\label{s011}
\eeq

Substituting (\ref{sp1})-(\ref{s011}) into (\ref{evol1}), (\ref{connect42}), for the contour ${\bm C_{-}}$, 
and noting that $|s_{ij}|^2 \equiv p_{j\rightarrow i}$ we find:

\beq
p_{1\rightarrow1} e^{\pi k} - p_{1 \rar 2} = e^{\pi k -\pi g^2/\beta}.
\label{cont1}
\eeq

Due to the unitarity of quantum mechanical evolution, the transition probability matrix is doubly  stochastic, which means that $p_{1\rightarrow1} + p_{1\rightarrow2} =1$. Combining this property with (\ref{cont1}) we find

\begin{eqnarray}
p_{1\rightarrow1}&=&p_{2\rightarrow 2}= \frac{e^{ -\pi g^2/\beta}+e^{-\pi k}}{1+e^{-\pi k}  }, \\
 p_{1 \rar 2}&=&p_{2 \rar 1}= \frac{1- e^{ -\pi g^2/\beta}}{1+e^{-\pi k}  }.
\label{prob11}
\end{eqnarray}
This result coincides with the solution of this model discussed in \cite{sinitsyn-13prl}. The case of $\beta<0$ can be worked out similarly but using either the rule for the contour ${\bm C_+}$ or applying the connection rule to the other state.

\subsection{Case 2: $\hat{A}=0$}
The most general, irreducible by elementary phase transformations, 2-state case with $\hat{A}=0$ reads:
\beq
i\frac{d}{dt} \left( \ba{c}
a\\
b
\ea \right)
= \left( \ba{cc}
\beta t & g/t\\
g/t & k/t 
\ea \right) \left( \ba{c}
a\\
b
\ea \right).
\label{hh2}
\eeq  

Equation~(\ref{hh2}) is symmetric under  reflection of time (i), which means that $\hat{S}_{-}=(\hat{S}_+)^{\dagger}$.
However, a small complication in comparison to the previous case follows from the fact that, at $t\rightarrow 0_{\pm}$, the diabatic states are not eigenstates of the Hamiltonian. 
Physically, it is expected that the evolution starts from some eigenstate of the Hamiltonian, and at  $t\rightarrow 0_{\pm}$, the Hamiltonian eigenstates coincide with
the eigenstates of the matrix $\hat{C}$.
Let $|+\ra$ and $|-\ra$ be the two eigenstates that correspond to eigenvalues
\beq
E_{\pm}= \frac{k\pm\sqrt{k^2+4g^2}}{2}
\label{cev1}
\eeq
of the matrix
\beq
\hat{C}=\left( \ba{rr}
0 & g\\
g & k
\ea \right).
\label{cc1}
\eeq
In the basis of states $|\pm\ra$, the evolution around the contour ${\bm C_0}$ in the upper half-plane has a  simple form:
\beq
 \hat{S}_0^{\rm up} =   \left( \ba{cc}
e^{-\pi E_{+}} & 0 \\
0 & e^{-\pi E_{-}}
\ea \right).
\label{s01}
\eeq

\begin{figure}
\scalebox{0.24}[0.24]{\includegraphics{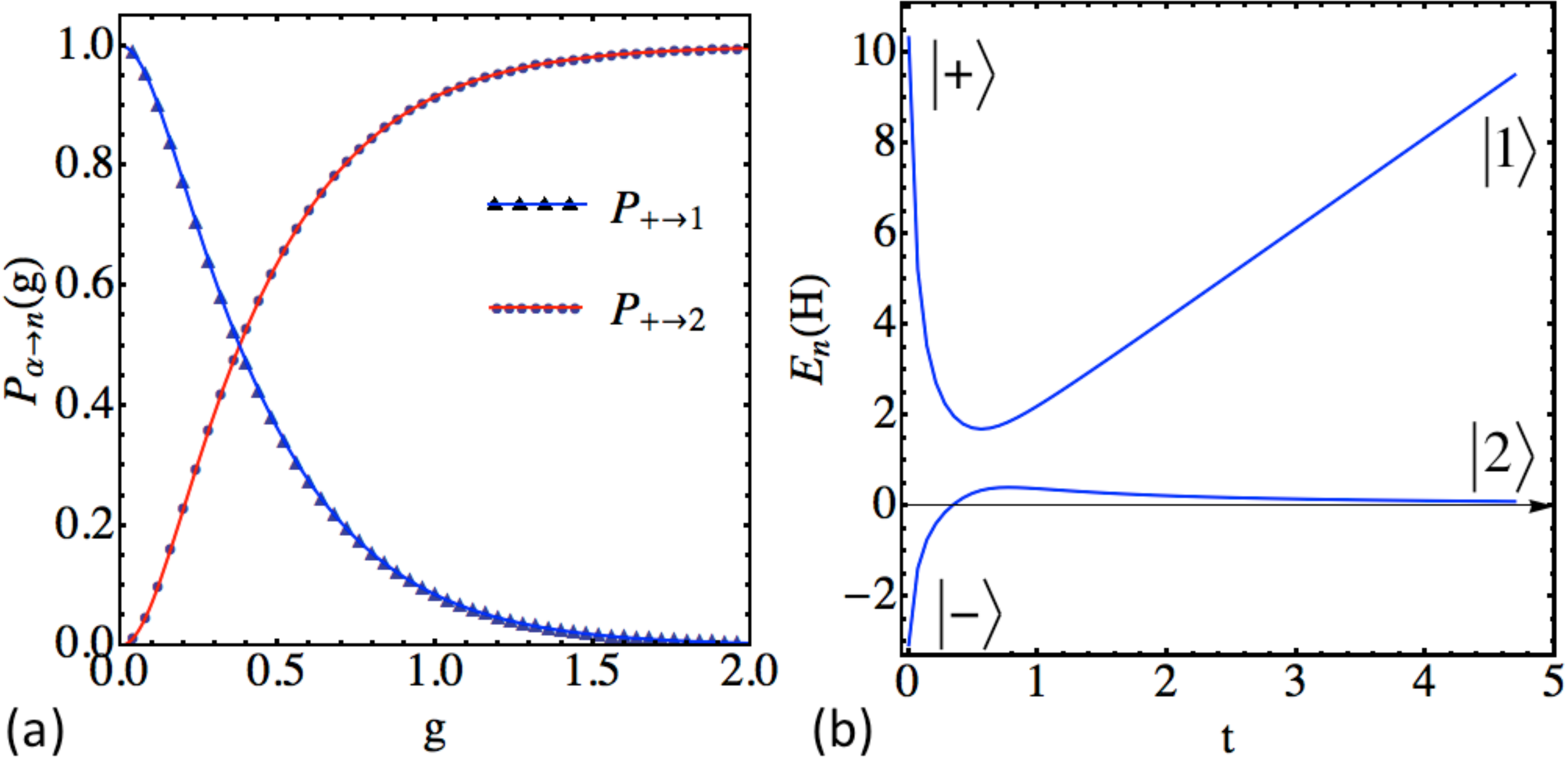}}
\hspace{-2mm}\vspace{-4mm}   
\caption{(a) Numerical test of Eq.~(\ref{prob11}). Solid curves are theoretical predictions and discrete points are numerical results. Parameters: $\beta=2$, $k=0.5$. Solution of Eq.~(\ref{mlz}) was obtained for the time interval $t \in (0.001,1000)$
with initial condition $|\psi\ra = |+\ra$. Details of the numerical program are discussed in supplementary file for Ref.~\cite{sinitsyn-13prl}.
 (b) Plot of eigenvalues of the Hamiltonian for the model (\ref{hh2}) as functions of time at $g=0.4$. At $t\rightarrow 0$, eigenstates coincide with states $|\pm \ra$, while at $t \rar +\infty$, eigenstatates approach the diabatic states $|1\ra$ and $|2\ra$.}
\label{test2}
\end{figure}

Scattering amplitudes from states $|\pm\ra$ and diabatic states are then well defined for evolution during $t\in (0_{+}, +\infty) $. Let 
$\hat{S}_+$ be such a scattering matrix with elements $s_{j \alpha}$, where $j=1,2$ and $\alpha=+,-$.  In combination with symmetry (i), the connection formula for the diabatic state $|1 \ra$, i.e. for the state that has the highest slope $\beta>0$, reads:

\beq
\left( \hat{S}_+  \hat{S}_0^{\rm up}  \hat{S}^{\dagger}_+  \right)_{11} =  1,
\label{con1}
\eeq
which can be written in terms of transition probabilities, $p_{\alpha \rar j}\equiv |s_{j\alpha}|^2$, as
\beq
e^{-\pi E_{+}} p_{+ \rar 1} +e^{-\pi E_{-}} p_{- \rar 1}=1.
\label{probs1}
\eeq
Using the unitarity constraint, $ p_{+ \rar 2}+ p_{- \rar 2} = 1$, we finally find: 
\begin{eqnarray}
\label{prob11}
p_{+\rightarrow 1}&=&p_{- \rightarrow 2}= \frac{e^{-\pi E_{-}} -1}{e^{-\pi E_{-}} -e^{-\pi E_{+}}  }, \\
 p_{- \rar 1}&=&p_{+ \rar 2}=\frac{1- e^{-\pi E_{+}}  }{e^{-\pi E_{-}} -e^{-\pi E_{+}}  }.
\end{eqnarray}
   In appendix B, we solve a multistate model that includes result (\ref{prob11}) at $k=0$ as a special case. In Fig.~\ref{test2}, we verify predictions of Eqs.~(\ref{prob11}) numerically by simulating the evolution (\ref{mlz}) with the Hamiltonian (\ref{hh2}).

\subsection{Case 3: $\hat{B}=0$}
Here we will explore two possibilities.  

${\bf (I)}$ First, we consider the evolution without diagonal elements of the matrix $\hat{C}$:
\beq
i\frac{d}{dt} \left( \ba{c}
a\\
b
\ea \right)
= \left( \ba{cc}
\epsilon & g/t\\
g/t & -\epsilon
\ea \right) \left( \ba{c}
a\\
b
\ea \right),
\label{hh3}
\eeq  
where $\epsilon>0$.

Since at $t\rightarrow +\infty$ the off-diagonal terms vanish, diabatic states become eigenstates of the Hamiltonian and one can define the scattering matrix from eigenstates 
of the corresponding matrix $\hat{C}$,
$$
|\pm\ra = \frac{|1\ra \pm |2\ra}{\sqrt{2}},
$$ 
to the diabatic states. 

The model (\ref{hh3}) has interesting property: since the matrix $\hat{B}$ is zero, all diabatic states can be considered as having both the highest and the lowest slope. Moreover, in addition to the previously used connection rules (\ref{connect4})-(\ref{connect41}) we have also the option to use the no-go rule (\ref{connect42}). Equation~({\ref{hh3}) is also symmetric under two, simultaneously applied, discrete operations: time reversal $t\rightarrow -t$ and a new discrete operation: 

(iv) exchange of indexes: $1\rar 2$, $2\rar 1$.
Consider the evolution matrix during time $t\in (0_+,+\infty)$ from states $|\pm \ra$ to diabatic states $|1\ra$, $|2\ra$:
 \beq
 \hat{S}_{+} =  \left( \ba{cc}
s_{1+} & s_{1-}\\
s_{2+} & s_{2-}
\ea \right).
\label{sp31}
\eeq
Under the symmetry (iv), state $|-\ra$ changes sign and states  $|1\ra$ and $|2\ra$ transfer into each other. Combining this fact with symmetry operations (i) and (iii), which were defined for previous models, we find the expression for the evolution matrix in  negative times  in terms of elements 
 $s_{j\alpha}$:
 
\beq
 \hat{S}_{-} =  \left( \ba{cc}
s_{2+}^* & s_{1+}^*\\
-s_{2-}^* & -s_{1-}^*
\ea \right).
\label{sp32}
\eeq
Finally, in the basis $|\pm \ra$ we have:

\beq
 \hat{S}_0^{\rm up} =   \left( \ba{cc}
e^{-\pi g} & 0 \\
0 & e^{\pi g}
\ea \right).
\label{s04}
\eeq
For evolution in the upper complex half-plane, it will be easiest to use the no-go rule (\ref{connect42}) that reads: 
\beq
\left( \hat{S}_+  \hat{S}_0^{\rm up}  \hat{S}_{-}  \right)_{12} = 0,
\label{con2}
\eeq
or explicity:
$$
|s_{1+}|^2 e^{-\pi g} - |s_{1-}|^2 e^{\pi g} = 0.
$$
Combining this with the definition of transition probabilities and the unitarity constraint
(i.e. that the matrix of transition probabilities is doubly stochastic), we finally obtain:
\begin{eqnarray}
p_{+\rightarrow 2}&=&p_{- \rightarrow 1}= \frac{1}{1+e^{2\pi g}  }, \\
 p_{- \rar 2}&=&p_{+ \rar 1}= \frac{1}{1+e^{-2\pi g}  }.
\label{prob13}
\end{eqnarray}

${\bf (II)}$ Next, we consider the most general case of evolution with zero matrix $\hat{B}$:
\beq
i\frac{d}{dt} \left( \ba{c}
a\\
b
\ea \right)
= \left( \ba{cc}
\epsilon +k/t & g/t\\
g/t & -\epsilon
\ea \right) \left( \ba{c}
a\\
b
\ea \right),
\label{hh4}
\eeq  
where $\epsilon>0$.
In such a case, there is no obvious symmetry that connects evolution at negative and positive times. However, the number of constraints that we can use includes two constraints (\ref{connect4})-(\ref{connect41}) 
that can be applied to each diabatic level. The latter is because each level can be considered having both the highest and the lowest slope in this model. In addition, there are two no-go constraints (\ref{connect42}) depending on whether we choose the  contour ${\bm C}$ in the upper or the lower half-plane. It turns out, that not all of those constraints are independent of each other but their number is sufficient to estimate transition probabilities, both for negative and positive time evolution.

As in Case 2, we will explore transition probabilities from states $|\pm \ra$ that are eigenstates of the matrix $\hat{C}$, with eigenvalues $E_{\pm}$ defined in (\ref{cev1}), to the diabatic states.

To derive transition probabilities, we will use the fact that any  $2\times2$ unitary matrix can be parametrized by three parameters, $p_1$, $\phi_1$ and $\theta_1$ as follows: 
\beq
 \hat{S}_+ =   \left( \ba{cc}
\sqrt{p_1} e^{i\phi_1} & \sqrt{1-p_1} e^{i\theta_1} \\
-\sqrt{1-p_1} e^{-i\theta_1} & \sqrt{p_1} e^{-i\phi_1} 
\ea \right).
\label{sp5}
\eeq
Similarly, we can parametrize scattering matrix for negative in time evolution: 
\beq
 \hat{S}_{-} =   \left( \ba{cc}
\sqrt{p_2} e^{i\phi_2} & \sqrt{1-p_2} e^{i\theta_2} \\
-\sqrt{1-p_2} e^{-i\theta_2} & \sqrt{p_2} e^{-i\phi_2} 
\ea \right),
\label{sm5}
\eeq
and the evolution around $t=0$  is given by 
\beq
 \hat{S}_0^{\rm up/dn} =   \left( \ba{cc}
e^{\mp \pi E_{+}} & 0 \\
0 & e^{\mp \pi E_{-}}
\ea \right).
\label{s05}
\eeq

The no-go rule applied to the contour in the upper half-plane gives 
\beq
\left( \hat{S}_+  \hat{S}_0^{\rm up}  \hat{S}_{-}  \right)_{12} = 0,
\label{con2}
\eeq
which in terms of the introduced parametrization reads:
\begin{eqnarray}
\nonumber e^{-\pi E_{-}} \sqrt{p_2(1-p_1)}  e^{-i(\phi_2-\theta_1)} + \\
 +e^{-\pi E_{+}}  \sqrt{p_1(1-p_2)}  e^{i(\phi_1+\theta_2)}=0.
\label{interm1}
\end{eqnarray}
Moving one of the terms in (\ref{interm1}) to the rhs and comparing absolute values, we obtain the relation between probabilities:
\beq
p_1(1-p_2)e^{-2\pi E_{+}} = p_2(1-p_1)E^{-2\pi E_{-}}.
\label{rel1}
\eeq

Another equation for probabilities is obtained by applying connection formulas (\ref{connect4})-(\ref{connect41}) to the 2nd diabatic state. 
Subtracting results for the upper and the lower contours from each other we find
\beq
\left( \hat{S}_+  \left[ \hat{S}_0^{\rm up} -\hat{S}_0^{\rm dn}  \right] \hat{S}_{-}  \right)_{22}=0,
\label{con2}
\eeq 
which leads to 
\beq
(1-p_1)(1-p_2)\left( e^{\pi E_{+} }-e^{-\pi E_{+}} \right)^2  =p_1p_2\left( e^{\pi E_{-}} -e^{-\pi E_{-}} \right)^2. 
\label{rel2}
\eeq
Solving (\ref{rel1}), and (\ref{rel2}) we find:
\beq
p_2=\frac{e^{2\pi E_{-} } ( e^{2\pi E_+}-1 ) } { e^{2\pi E_+}-e^{2\pi E_{-} } },
\label{rel22}
\eeq
\beq
p_1=\frac{e^{2\pi E_+}-1} {e^{2\pi E_+}-e^{2\pi E_{-}} }.
\label{rel3}
\eeq
Note that $p_1\ne p_2$, i.e. we were able to find nontrivial transition probabilities  simultaneously for the negative and the positive evolution time intervals. Finally, with unitarity constraints, we can identify transition probabilities for positive time:
\beq
p_{+\rightarrow 1}= p_{-\rightarrow 2} =p_1, \quad p_{+\rightarrow 2}= p_{-\rightarrow 1}=1-p_1. 
\label{prob22}
\eeq
This example shows that it is not always necessary to use  discrete symmetries in order to  obtain interesting results with connection formulas.

\section{Multistate LZC systems}
The two-state systems that were solved analytically in previous section can, in principle, be solved by other means.  For example, all of them can be reduced to the well-understood confluent hypergeometric equation. In contrast, much less is known about how to solve systems with $N>2$. Therefore, this section
contains most important results that demonstrate how connection formulas can make a nontrivial insight in the behavior of transition probabilities in multistate models.


\subsection{Special State Model}
Consider a model in which arbitrary number $N$ of states interact with  a single special state, to which we will give zero index, and

(i) matrix $\hat{B}$ is not degenerate;

(ii) matrix $\hat{A}$ contains nonzero elements only as couplings of the special state to the other states, i.e. $A_{i0}=A_{0i} = g_i$ and all other elements of $\hat{A}$ are zero;
 
(iii) matrix $\hat{C}$ has arbitrary nonzero elements except couplings to the special state, i.e. $C_{i0}=C_{0i}=0$.

The Hamiltonian of such a system in the diabatic basis has the form:

\beq
 \hat{H}(t) =   \left( \ba{ccccc}
k_0/t & g_1& g_2& \ldots & g_N \\
g_1 &\beta_1 t +k_1/t & g_{12}/t & \ldots & g_{1N}/t\\
g_2& g_{12}/t & \beta_2 t +k_2/t &\ldots& \ldots \\
\vdots& \vdots & \vdots &\ddots & \vdots
\ea \right).
\label{ham6}
\eeq

Let 
$s_{j\alpha}$, where $\alpha, j \in (0,\ldots N)$, be the transition amplitude from the $\alpha$-th eigenstate of $\hat{C}$ to the $j$-th diabatic state for positive in time evolution from $t\rar 0_+$ to $t \rar +\infty$, i.e. 
\beq
\hat{S}_+ =   \left( \ba{cccc}
s_{00} & s_{01} &  \ldots & s_{0N} \\
s_{10} &s_{11} & \ldots & \vdots\\
\vdots& \vdots & \ddots & \vdots
\ea \right).
\label{sp61}
\eeq
Evolution equation (\ref{mlz}) with the Hamiltonian (\ref{ham6}) is symmetric under the time reflection, $t \rar -t$, followed by  the change of the sign of the amplitude of the special state.  In turn, this means that 
the scattering matrix for the evolution from  $t\rar -\infty$ to $t \rar 0_{-}$ has the form: 
\beq
\hat{S}_- =   \left( \ba{ccccc}
s^*_{00} & -s^*_{10} &  \ldots &\ldots  & -s^*_{N0} \\
-s_{01}^* & s^*_{11} &s^*_{21} & \ldots & \vdots\\
-s_{02}^* & s^*_{12} &s^*_{22} & \ldots & \vdots\\
\vdots& \vdots &\vdots & \ddots & \vdots
\ea \right).
\label{sm61}
\eeq

Suppose, first, that the state $|0 \ra$ has the highest/lowest slope. Then in the basis of eigenstates of $\hat{C}$, we have:
 $\hat{S}_0^{\rm up} = {\rm diag} \{e^{\mp \pi k_0}, e^{\mp  \pi E_1} ,\ldots e^{\mp \pi E_N} \}$, where $E_i$ are eigenvalues of the matrix $\hat{C}$, and the choice of 
 $-/+$ depends on whether the state has the highest/lowest slope, which in turn  determines whether the contour ${\bm C_0}$ should be in the upper or in the lower complex half-plane. Substituting this and (\ref{sp61})-(\ref{sm61}) into (\ref{connect4})-(\ref{connect41}) we obtain:
 
 \beq
 e^{\mp \pi k_0} p_{0\rightarrow 0} -\sum_{i=1}^N  e^{ \mp \pi E_i}   p_{i\rightarrow 0} =  e^{\pi ( \mp k_0 - \sum_{i=1}^N g_i^2/|\beta_i|)}.
 \label{p61}
 \eeq
If, instead, a level with $n\ne 0$ has the highest/lowest slope, then (\ref{connect4})-(\ref{connect41}) lead to the constraint:
 \beq
 -e^{\mp \pi k_0} p_{0\rightarrow n} +\sum_{i=1}^N  e^{\mp \pi E_i}   p_{i\rightarrow n} =  e^{\pi (\mp k_n -g_n^2/|\beta_n|)}.
 \label{p63}
 \eeq

Although Eqs.~(\ref{p61})-(\ref{p63}) do not determine a particular element of the transition probability matrix, they represent exact nonperturbative constraints that reduce the number of unknown independent parameters of the transition probability matrix. In special cases  these equations can be used to derive specific probabilities.  Consider, e.g., the situation in which only one element of the matrix $\hat{C}$ is nonzero, i.e.  $C_{00}=k_0$. 
In such a case, $E_i=0$ for $i=1,\ldots N$. Using the unitarity condition, $\sum_{i=0}^N   p_{i\rightarrow 0} =1$, Eqs.~(\ref{p61})-(\ref{p63}) tell that if  the level $0$ is extremal then 
\beq
p_{0\rightarrow 0} = \frac{1+e^{\pi ( \mp k_0 - \sum_{i=1}^N g_i^2/|\beta_i|)}}{1+ e^{\mp \pi k_0}},
\label{p62}
\eeq
and if a level with $n\ne 0$ is extremal then 
 \beq
 p_{0\rightarrow n} =  \frac{1-e^{-\pi g_n^2/|\beta_n|}}{ 1+ e^{\mp \pi k_0}} ,
 \label{p64}
 \eeq
where $-/+$ corresponds to the situation in which a given level has the highest/lowest slope. Probabilities (\ref{p62})-(\ref{p64}) coincide with their values known from the exact solution of this model \cite{sinitsyn-13prl}.

\subsection{Model with two special states }

Consider now a generalization of the previous model in which two states, with indexes $0$ and $0'$, equally interact with other states. Diabatic energies of those states are separated by a finite distance:
$A_{00} = -A_{0'0'}=\epsilon$ and those states are allowed to interact with each other with a decaying coupling: $C_{00'}=C_{0'0}=g/t$. The Hamiltonian of this system can be written in the following matrix form: 

\beq
 \hat{H}(t) =   \left( \ba{ccccc}
\epsilon & g/t& g_1& \ldots & g_N \\
g/t & -\epsilon& g_1 & \ldots& g_N \\
g_1& g_1& \beta_1 t +k_1/t & g_{12}/t & \ldots \\
g_2& g_2& g_{12}/t & \beta_2 t +k_2/t &  \ldots \\
\vdots& \vdots & \vdots &\ddots & \vdots
\ea \right).
\label{ham777}
\eeq
Matrix $\hat{C}$ has two eigenvectors that can be written explicitly:
 
\beq
|\pm \ra = \frac{1}{\sqrt{2}} \left( |0 \ra \pm |0'\ra \right),
\label{pm7}
\eeq 
with eigenvalues $E_{\pm} = \pm g$. For other eigenvalues of $\hat{C}$ we will use notation of the previous model. For example, 
 $\hat{S}_0^{\rm up/dn} = {\rm diag} \{e^{\mp  \pi E_+},e^{\mp  \pi E_{-}}, e^{\mp  \pi E_1} ,\ldots e^{\mp \pi E_N} \}$.

Evolution equation (\ref{mlz}) with the Hamiltonian (\ref{ham777}) is symmetric under simultaneous time reversal, change of sign of special state amplitudes, and exchange of their indexes: $0\rar 0'$ and $0' \rar 0$.
The latter operation leaves the state $|+\ra$ invariant and changes the sign of $|-\ra$. 
Consequently, if the  scattering matrix for positive time has the form
\beq
\hat{S}_+ =   \left( \ba{ccccc}
s_{0+} & s_{0-} & s_{01} & \ldots & s_{0N} \\
s_{0'+} &s_{0'-}& s_{0'1}  & \ldots & \vdots\\
s_{1+} &s_{1-}& s_{11}  & \ldots & \vdots\\
\vdots& \vdots & \ldots &\ddots & \vdots
\ea \right),
\label{sp71}
\eeq
then the scattering matrix for the negative time evolution has the form:
\beq
\hat{S}_{-} =   \left( \ba{ccccc}
s_{0'+}^* & s^*_{0+} & -s_{1+}^* & \ldots & -s_{N+}^* \\
-s_{0'-}^* & -s_{0-}^*& s_{1-}^*  & \ldots & \vdots\\
-s_{0'1}^*&-s_{01}^*& s_{11}^*  & \ldots & \vdots\\
\vdots& \vdots & \ldots &\ddots & \vdots
\ea \right).
\label{sm71}
\eeq
Suppose that the special levels have the highest slope. Then the transition from the state $|0' \ra$ to the state $|0 \ra$ is counterintuitive. The No-Go Rule then produces a simple relation for transition probabilities:

\beq
e^{-\pi g} p_{+ \rar 0} -e^{\pi g} p_{- \rar 0}  - \sum_{i=1}^N e^{-\pi E_i} p_{i \rar 0} = 0. 
\label{nogo711}
\eeq
If, in this situation, the level with index $n\ne \{0,0' \}$ has the lowest slope then the connection rule (\ref{connect41}) gives:
\beq
-e^{\pi g} p_{+ \rar n} + e^{-\pi g} p_{- \rar n} + \sum_{i=1}^N e^{\pi E_i} p_{i \rar n} = e^{\pi (k_n-2g_n^2/|\beta_n|)}. 
\label{con712}
\eeq

For example, consider a special case: $C_{ii}=-g$ and $C_{ij}=0$, for $i,j=1,\ldots N$.
In such a case, $E_i=k_i=-g$, and after using the doubly stochastic character of the transition probability matrix, Eqs.~(\ref{nogo711})-(\ref{con712}) produce simple results:
\beq
 p_{+ \rar 0} = \frac{1}{1+e^{-2\pi g}}, \quad p_{+ \rar n} = \frac{1- e^{-2\pi g_n^2/|\beta_n|}} {1+e^{2\pi g}}. 
\label{con722}
\eeq

\begin{figure}
\scalebox{0.24}[0.24]{\includegraphics{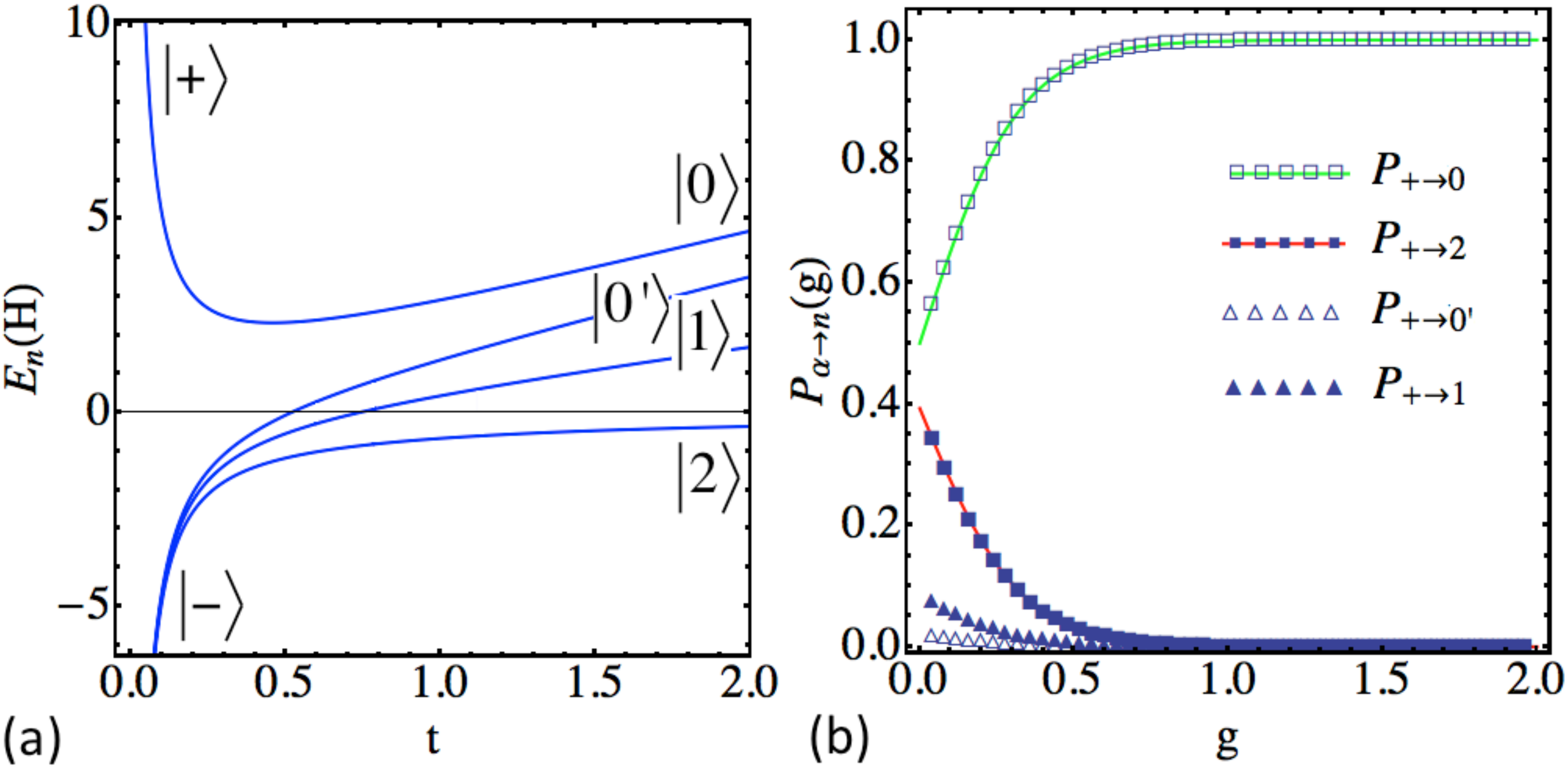}}
\hspace{-2mm}\vspace{-4mm}   
\caption{(a)  Plot of eigenvalues of the Hamiltonian  (\ref{htest}). (b) Numerical test of Eq.~(\ref{con722}). Parameters for numerical simulations: $g_1=0.5$, $g_2=0.7$, $\epsilon=0.5$. Evolution time interval is $t \in (0.001,1000)$. Solid curves are theoretical predictions and discrete points are numerical results. }
\label{test-fig41}
\end{figure}

For a numerical check, we consider the latter model with $N=4$ and the Hamiltonian
\beq
\hat{H} =   \left( \ba{cccc}
\epsilon +2 t& g/t & g_1 & g_2 \\
g/t &-\epsilon + 2 t& g_1 & g_2 \\
g_1 &g_1&t-g/t  & 0\\
g_2& g_2 & 0 & -g/t 
\ea \right).
\label{htest}
\eeq
In Fig.~\ref{test-fig41}, we illustrate eigenvalues of this Hamiltonain as functions of $t$ and compare Eq.~(\ref{con722}) with results of numerical simulations, which are found to be in perfect agreement with each other.

\subsection{Chain models}
There are numerous physical applications, in which diabatic states are coupled by a constant coupling in a chain-like fashion with all other elements of matrix $\hat{A}$ being zero \cite{chain1}.  
In such a case,  Eq.~(\ref{mlz}) transforms into the set of coupled equations:
\beq
i \dot{\psi}_n = (k_n/t + \beta_n t ) \psi_n+g_n \psi_{n+1} + g_{n-1}\psi_{n-1}, 
\label{chain1}
\eeq
where $n=1,\ldots, N$, and where we define $g_0=g_N=0$.  For evolution (\ref{chain1}), diabatic states coincide with eigenstates both at $t\rightarrow 0$ and at $t\rar \infty$.
Equation~(\ref{chain1}) is symmetric under simultaneous time reversal $t\rightarrow -t$ and the change of the sign of the amplitudes with even indexes. Therefore, the scattering matrix for negative
time is written in terms of matrix elements $s_{ij}$, $i,j =1,\ldots, N$, for positive time as  
\beq
\hat{S}_{-} =   \left( \ba{cccc}
s_{11}^* & -s^*_{21} & s_{31}^* & \ldots  \\
-s_{12}^* & s_{22}^*& -s_{32}^*  & \vdots\\
s_{13}^*&-s_{23}^*& s_{33}^*   & \vdots\\
\vdots& \vdots & \ldots &\ddots 
\ea \right),
\label{smc1}
\eeq

Consider the case when a state with index $n$ has the highest/lowest slope. Then if $n$ is odd, Eqs.~(\ref{connect4})-(\ref{connect41}) return:
\beq
\sum_{i=1}^{N} (-1)^{i+1} p_{i \rar n} e^{\mp \pi k_i} = e^{-\pi \left(\pm  k_n +\frac{g_{n}^2}{|\beta_n-\beta_{n+1}|} + \frac{g_{n-1}^2}{|\beta_n-\beta_{n-1}|}\right)}.
\label{pch1}
\eeq
If $n$ is even then 
\beq
\sum_{i=1}^{N} (-1)^{i} p_{i \rar n} e^{\mp \pi k_i} = e^{-\pi \left( \pm k_n +\frac{g_{n}^2}{|\beta_n-\beta_{n+1}|} + \frac{g_{n-1}^2}{|\beta_n-\beta_{n-1}|}\right)},
\label{pch2}
\eeq
where $-/+$ corresponds to the highest/lowest slope of the extremal level.

Here we note also that the same symmetry and, hence, the form of the scattering matrix (\ref{smc1}) is obtained if we generalize the chain model to  include  (a) constant couplings between states with arbitrary even and odd indexes and (b) decaying with time couplings between states of the same index parity.
 Equasions~(\ref{pch1})-(\ref{pch2}) are straightforward to generalize to these situations.

\subsection{Models with $\hat{A}=0$}

Equation~(\ref{mlz}), in the case of $\hat{A}=0$, arbitrary $\hat{C}$, and non-degenerate $\hat{B}$,  is symmetric under reflection $t\rightarrow -t$. 
Let $E_{\alpha}$, $\alpha=1,\ldots, N$ be the eigenvalues of the matrix $\hat{C}$. Then for the extremal state with index $n$, (\ref{connect4}) gives:

\beq
\sum_{\alpha=1}^N e^{\mp \pi E_\alpha}p_{\alpha \rar n} = e^{\mp \pi k_n},
\label{pr8}
\eeq 
where $+/-$ corresponds to the highest/lowest slope.

As an example, consider the case when all levels are coupled to each other according to the rule 
\beq
C_{ij} = q_iq_j, \quad i, j=1,\ldots, N,
\label{all1}
\eeq
with $N$ independent constants $q_i$.
In such a case, matrix $\hat{C}$ has all zero eigenvalues except the one that corresponds to the state  vector 
\beq
|+\ra = \frac{1}{\sqrt{E_+}} \left( q_1|1\ra + \ldots + q_N |N\ra  \right),
\label{all2}
\eeq
with a single nonzero eigenvalue $E_+=\sum_{i=1}^N q_i^2$. Note also that $k_n=q_n^2$. Substituting this into (\ref{pr8}) and using the unitarity condition we find
\beq
p_{+\rar n}= \frac{1-e^{\mp \pi q_n^2}}{1-e^{\mp \pi \sum_{i=1}^N q_i^2}}.
\label{all3}
\eeq
In Appendix B, we show that, for the latter model, one can derive explicit expressions for transition probabilities from the state $|+ \ra$ to any other state by an alternative approach. The final result is in perfect agreement with (\ref{all3}).

\subsection{Models with $\hat{B}=0$}
Here we will explore two specific 3-state systems. 

\subsubsection{Case-1: Equal Coupling Model} 
Consider a 3-state model with the Hamiltonian
\beq
 \hat{H}(t) =   \left( \ba{ccc}
\epsilon & g/t& g/t  \\
g/t& 0 & g/t  \\
g/t& g/t & -\epsilon 
\ea \right),
\label{ham7}
\eeq
which eigenvalues as functions of time are shown in Fig.~\ref{test31}(a).
Matrix $\hat{C}$ has an eigenvalue $E_+=2g$ with an eigenvector 
\beq
|+\ra = \frac{1}{\sqrt{3}} \left( |1\ra + |2 \ra + |3 \ra \right), 
\label{ev81}
\eeq
and two degenerate eigenvalues $E_{0}=E_{0'}=-g$ with eigenvectors
\beq
|0\ra = \frac{|1\ra - |3 \ra }{\sqrt{2}},  \quad |0'\ra = \frac{|1\ra - 2|2 \ra  + |3 \ra}{\sqrt{6}}.
\label{ev71}
\eeq

\begin{figure}
\scalebox{0.24}[0.24]{\includegraphics{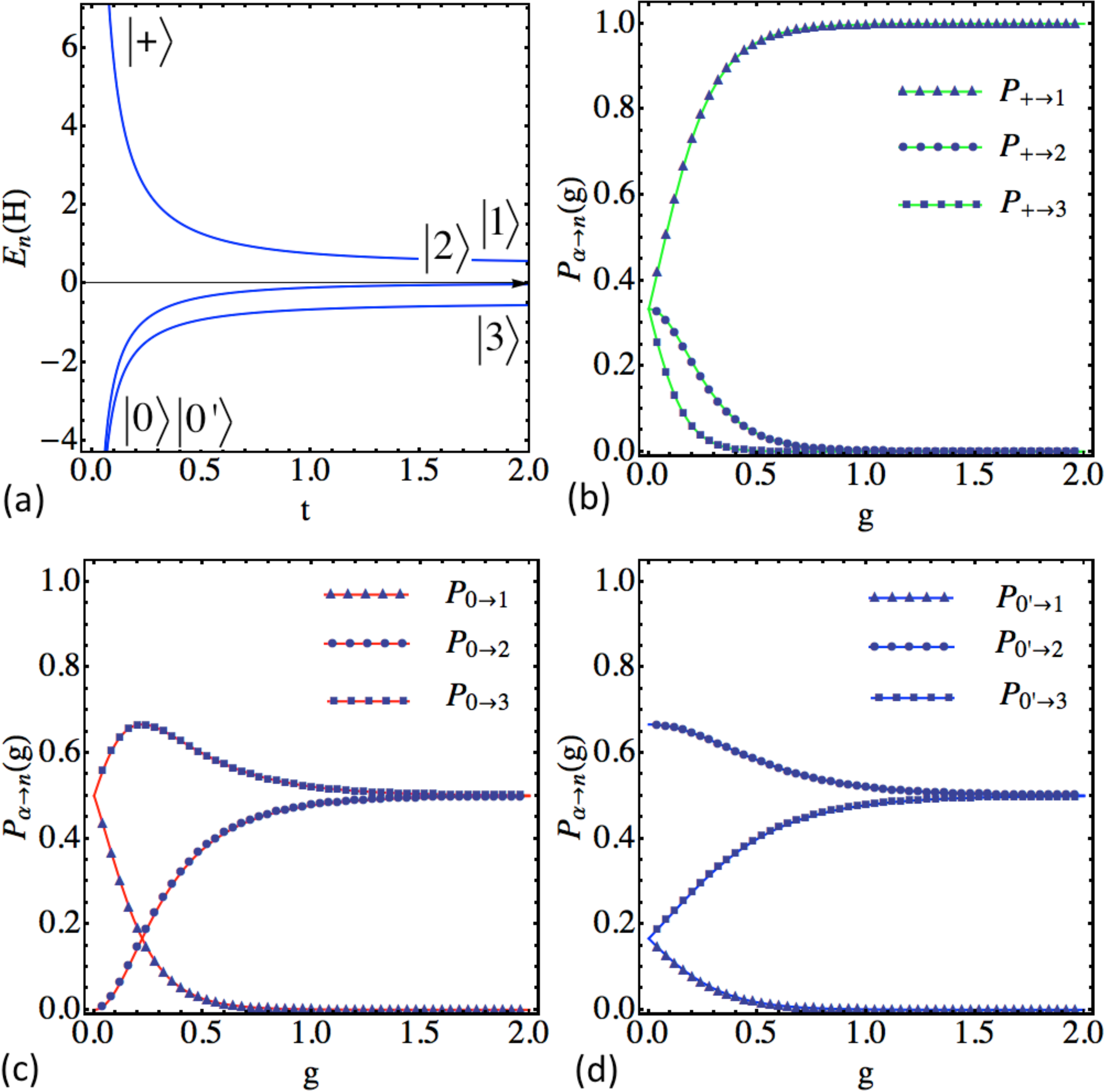}}
\hspace{-2mm}\vspace{-4mm}   
\caption{(a)  Plot of eigenvalues of the Hamiltonian  (\ref{ham7}) as functions of time at $g=0.3$, $\epsilon=0.5$. (b-d) Numerical test of Eqs.~(\ref{pr71}), (\ref{dual3}) at $\epsilon=0.5$ for different couplings $g$. Evolution time interval is $t \in (0.001,1000)$. Solid curves are theoretical predictions and discrete points are results of the numerical solution of the evolution equation with the Hamiltonian (\ref{ham7}). }
\label{test31}
\end{figure}

Equation~(\ref{mlz}) with the Hamiltonian (\ref{ham7}) is symmetric under the time reversal and, simultaneously, exchange of indexes $1 \rar 3$ and $3 \rar 1$. 
Note that the index exchange leaves states $|+\ra$ and $|0'\ra$ invariant  but changes the sign of $|0\ra$. 

Let the scattering matrix for the positive time evolution have the form
\beq
\hat{S}_+ =   \left( \ba{ccc}
s_{1+} & s_{10} & s_{10'}  \\
s_{2+}& s_{20}  & s_{20'}  \\
s_{3+}& s_{30} & s_{30'} 
\ea \right),
\label{s71}
\eeq
then the negative time scattering matrix reads:
\beq
\hat{S}_{-} =   \left( \ba{ccc}
s_{3+}^* &  s_{2+}^*  & s_{1+}^*  \\
-s_{30}^*& -s_{20}^*  & -s_{10}^*  \\
s_{30'}^* & s_{20'}^* & s_{10'}^* 
\ea \right),
\label{s72} 
\eeq
and in the basis of states (\ref{ev81})-(\ref{ev71})
\beq
\hat{S}_{0}^{\rm up/dn} = {\rm diag}\{e^{\mp \pi E_+}, e^{\mp \pi E_{0}}, e^{\mp \pi E_{0'}} \}.   
\label{s73}
\eeq

Here we note again that the case with $\hat{B}=0$ is, in some sense, unusual: all its states can be simultaneously considered as having both the highest and the lowest slope and all off-diagonal transitions can be considered counterintuitive, although depending on whether levels are considered having the highest or the lowest slope.
Applying the No-Go Rule to transitions between levels $3$ and $1$, we find constraints on probabilities:
\beq
p_{+\rar 1} e^{-2\pi g} - p_{0 \rar 1} e^{\pi g} +p_{0'\rar 1} e^{\pi g}=0,
 \label{nogo71}
\eeq
\beq
p_{+\rar 3} e^{2\pi g} - p_{0 \rar 3} e^{-\pi g} +p_{0'\rar 3} e^{-\pi g}=0,
 \label{nogo72}
\eeq
and applying the connection formulas (\ref{connect4})-(\ref{connect41}) to the state $|2 \ra$ we find 
\beq
p_{+ \rar 2} e^{\pm 2 \pi g} - p_{0 \rar 2} e^{\mp \pi g} + p_{0' \rar 2} e^{\mp \pi g} =1.
\label{con71}
\eeq
In combination with the unitarity rule: $p_{+ \rar 2} + p_{0 \rar 2} + p_{0' \rar 2}  =1$, Eq.~(\ref{con71}) leads to explicit expressions for transitions to the 2nd state:
\begin{widetext}
\beq
p_{+ \rar 2}=\frac{1}{1+2 \cosh (2\pi g)}, \quad  p_{0 \rar 2}=\frac{\cosh (\pi g) -1}{2 \cosh (\pi g)-1}, \quad   p_{0' \rar 2}=\frac{\cosh (\pi g) +1}{2 \cosh (\pi g)+1}.
\label{pr71}
\eeq
\end{widetext}

At a first view, it seems that  rules (\ref{nogo71})-(\ref{nogo72}) are insufficient to determine the remaining unknown elements of the transition probability matrix,
while the interpretation of the unused constraints in terms of the probabilities seems obscure. However, below, we will show that the model with $\hat{B}=0$ contains one extra useful property that, in our case, produces new simple constraints on transition probabilities.

\subsubsection{Duality between $\hat{B}=0$ and $\hat{A}=0$ models}
 
Consider arbitrary model (\ref{mlz}) with $\hat{A}=0$:

 \begin{equation}
i\frac{d\psi}{d t} = \left( \hat{B}t +\frac{ \hat{C}}{t} \right)\psi.
\label{mlz-a0}
\end{equation} 
For strictly positive time, $t>0$, we can make a change of variables: $t^2/2=\tau$. Transition from $t$ to $\tau$ does not change the scattering matrix for evolution during
$t \in (0_+, + \infty)$. Using that $d/dt = t d/d \tau$, we then find:  
 \begin{equation}
i\frac{d\psi}{d \tau} = \left( \hat{B} +\frac{ \hat{C}}{2 \tau} \right)\psi,
\label{mlz-a01}
\end{equation} 
i.e. the change of variables that does not affect transition probability matrix for $t>0$ transforms the model with ($\hat{A}=0$) into the model with ($\hat{B}=0$) but 
with elements of the matrix $\hat{C}$ rescaled by a factor $1/2$.

This means that we can apply Eq.~(\ref{pr8}) to the levels having the extremal (largest or lowest) value of the parameter $\epsilon_i$ in 
any model  of the $\hat{B}=0$ type. In particular, application of  this rule to the extremal levels of the model (\ref{ham7}) gives:
\beq
e^{-4\pi g} p_{+ \rar 1} + e^{2\pi g} p_{0 \rar 1} +e^{2\pi g} p_{0' \rar 1}=1,
\label{dual1}
\eeq
\beq
e^{4\pi g} p_{+ \rar 3} + e^{-2\pi g} p_{0 \rar 3} +e^{-2\pi g} p_{0' \rar 3}=1.
\label{dual2}
\eeq
In fact, one of expressions (\ref{dual1})-(\ref{dual2}) is redundant, as only one of them is sufficient to reconstruct all remaining transition probabilities:
\begin{widetext}
\begin{eqnarray}
\nonumber p_{+ \rar 1} &=&\frac{e^{4\pi g}}{1+e^{2\pi g}+e^{4\pi g}} , \quad p_{0 \rar 1} = \frac{1}{2-2e^{\pi g} +2e^{2\pi g} }, \quad 
p_{0' \rar 1} =\frac{1}{2(1+e^{\pi g} + e^{2\pi g})}, \\
\quad p _{+ \rar 3} &=&  \frac{1}{1+e^{2\pi g}+e^{4\pi g}}, \quad 
 p_{0 \rar 3} = \frac{e^{\pi g}}{4 \cosh (\pi g) -2}, \quad \,\,\, p_{0' \rar 3} =\frac{e^{\pi g}}{4 \cosh (\pi g) +2}.
\label{dual3}
\end{eqnarray}
\end{widetext}
In  Fig.~\ref{test31}(b-d) we provide numerical test of (\ref{pr71}) and (\ref{dual3}) that shows perfect agreement of theory and numerics. 

\subsubsection{Case-2: Chain model with decaying couplings} 
Consider another example of a 3-state model with the Hamiltonian
\beq
 \hat{H}(t) =   \left( \ba{ccc}
\epsilon & g/t& 0  \\
g/t& k/t & g/t  \\
0 & g/t & -\epsilon 
\ea \right),
\label{ham75}
\eeq
which eigenvalues, as functions of $t$, are shown in Fig.~\ref{test32}(a). 
Corresponding matrix $\hat{C}$ has one  eigenstate
\beq
|0 \ra = \frac{1}{\sqrt{2}} \left(|1 \ra - |3 \ra \right)
\label{zero71}
\eeq
 that corresponds to the zero eigenvalue, and two eigenstates, $|+\ra$ and $|-\ra$, that correspond to eigenvalues
 \beq
 E_{\pm} = \frac{1}{2} \left( k\pm \sqrt{k^2+8 g^2} \right).
 \label{nonzero71}
 \eeq
 One can check that under exchange of indexes $1\rar 3$ and $3 \rar 1$, eigenstate $|0 \ra$ changes sign, while eigenstates $|\pm \ra$ remain invariant.
The No-Go Rule then produces constraints:
\beq
p_{+\rar 1} e^{-\pi E_{+}} - p_{0 \rar 1} +p_{- \rar 1} e^{-\pi E_{-}}=0,
 \label{nogo76}
\eeq
\beq
p_{+\rar 3} e^{\pi E_{+}} - p_{0 \rar 3} +p_{- \rar 3} e^{\pi E_{-}}=0.
 \label{nogo77}
\eeq
The connection formulas (\ref{connect4})-(\ref{connect42}) applied to the diabatic state $|2 \ra$ produces 
\beq
p_{+ \rar 2} e^{\mp \pi E_{+}} - p_{0 \rar 2}  + p_{- \rar 2} e^{\mp \pi E_{-}} =e^{\mp \pi k}, 
\label{con78}
\eeq
and the duality produces an additional constraint: 
\beq
p_{+ \rar 1} e^{- 2\pi E_{+}} + p_{0 \rar 1}  + p_{- \rar 1} e^{-2 \pi E_{-}} =1.
\label{con79}
\eeq

Altogether, Eqs.~(\ref{nogo76})-(\ref{con79}) and the doubly stochastic character of the transition probability matrix produce the 
set of transition probabilities:
\begin{widetext}
\begin{eqnarray}
  \label{dual5}
\nonumber p_{+ \rar 2} &=&\frac{e^{-\pi E_{-}}(1+e^{\pi k}) ( e^{\pi k} -e^{\pi E_{-}})  }{ (1+e^{\pi E_+}) (e^{\pi E_+} - e^{\pi E_{-}}) }, \quad  
 p_{0 \rar 2} = \frac{e^{\pi E_+ } +e^{\pi E_{-}} -1 - e^{\pi k} } {1+e^{\pi E_+ } +e^{\pi E_{-}} +e^{\pi k} }, \quad   
 p_{- \rar 2} = \frac{e^{-\pi E_{+} }(1+e^{\pi k}) ( e^{\pi E_{+}}- e^{\pi k} )  }{ (1+e^{\pi E_-}) (e^{\pi E_+} - e^{\pi E_{-}}) }, \quad \\
 \nonumber \\
 &&\quad \quad  \quad \quad p_{- \rar 3} =e^{-2\pi E_{-}} p_{- \rar 1},  \quad \quad  p_{+ \rar 3} =  e^{-2 \pi E_{+}} p_{+ \rar 1},\\
  \nonumber \\
 \nonumber  p_{+\rar 1} &=&  \frac{e^{2\pi E_{+} } (1-e^{\pi E_{-} })} {(e^{\pi E_{+}} +1)(e^{\pi E_{+}}-e^{\pi E_{-}}) }, \quad
 p_{-\rar 1} =  \frac{e^{2\pi E_{-} } (e^{\pi E_{+} }-1)} {(e^{\pi E_{-}} +1)(e^{\pi E_{+}}-e^{\pi E_{-}}) }, \quad
 p_{0\rar 1} =p_{0\rar 3}=  \frac{1+e^{\pi k } } {1+e^{\pi E_+ } +e^{\pi E_{-}} +e^{\pi k} }.
\end{eqnarray}
\end{widetext}

\begin{figure}
\scalebox{0.23}[0.23]{\includegraphics{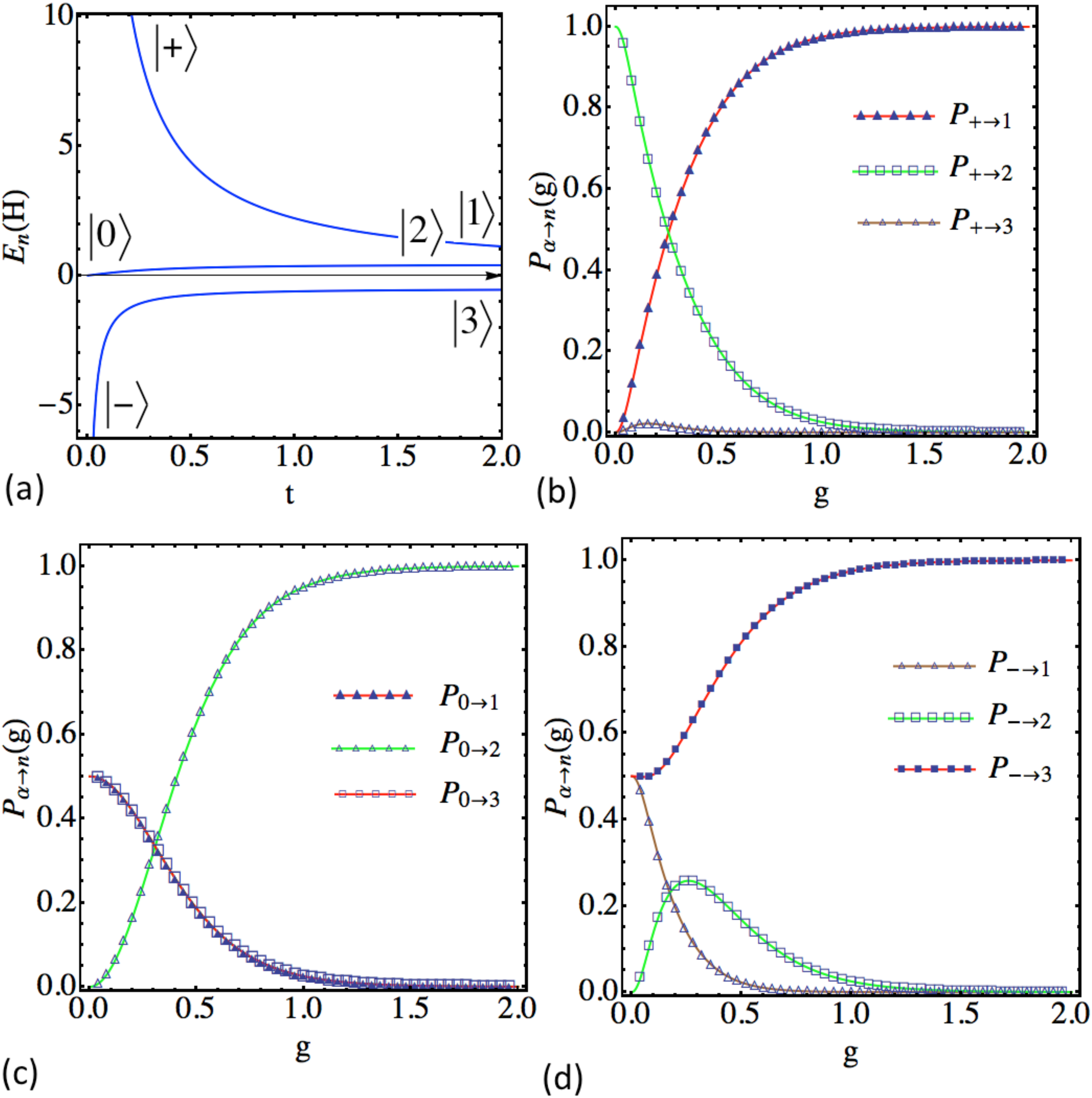}}
\hspace{-2mm}\vspace{-4mm}   
\caption{(a)  Plot of eigenvalues of the Hamiltonian  (\ref{ham75}) as functions of time at $g=\epsilon=0.5$, $k=2$. (b-d) Numerical test of Eqs.~(\ref{dual5}) at $\epsilon=0.5$, $k=0.3$ for different couplings $g$. Evolution time interval is $t \in (0.001,1000)$. Solid curves are theoretical predictions and discrete points are results of the numerical solution of the evolution equation with the Hamiltonian (\ref{ham75}). }
\label{test32}
\end{figure}

In Fig.~\ref{test32}(b-d) we compare theoretical predictions (\ref{dual5}) with transition probabilities obtained with numerical simulations and find perfect agreement between them.

\section{Discussion and Conclusion}



In this work, we demonstrated that the  absence of the Stokes phenomenon is the property of solutions in a large class of LZC-systems (\ref{mlz}). 
For any such a model, it is possible to obtain exact nontrivial constraints on the elements of the scattering matrix. Generally, those constraints  do not show a simple interpretation in terms of transition 
probabilities for evolution from   $t \rightarrow 0_+$ to $t\rightarrow +\infty$. However, there is quite a large subclass of LZC systems that contain additional discrete symmetries that eventually lead to simple linear constraints on elements of the transition probability matrix and, sometimes, even to analytical expressions for probabilities of particular transitions.

Certainly, it is likely that examples found in this article do not exhaust the set of tricks that can be applied to derive new interesting solutions of LZC models. So
the future progress in this direction is expected. 
Here we would like to point to  alternative research directions, which have not been explored in the present article.

 First, we note that the proof of the absence of the Stokes phenomenon can be applied to even larger class of systems.
For example, one can consider
the {\it Generalized LZC Model} with the Hamiltonian
 \begin{equation}
\hat{H}=\hat{A} + \hat{B}t + \sum_{i=0}^L \frac{\hat{C}_i }{t-t_i}
\label{LZCC2}
\end{equation}
that has similar,  to the LZC model, behavior of asymptotic solutions near points $t_i$ and $t\rightarrow \infty$.
In such a case, a contour at $R\rightarrow \infty$ that connects asymptotics at $t \rightarrow \pm \infty$ can be continuously deformed to lay along the real axis except the points $t_i$ that it should encircle. 
 It would be interesting to find out whether it is possible to derive useful constraints on transition probabilities for models of the type (\ref{LZCC2}) by imposing additional symmetries of the Hamiltonian, as we did in the present work for the LZC model.

The second observation that can be useful  is that  even if only one of the amplitudes is completely known along a contour then all other solutions can be found as integrals of this known amplitude. 
For example, an arbitrary evolution of a two-state system with time-dependent coefficients can be written in the following form:
\begin{eqnarray}
\nonumber i\dot{a}&=&e(t) a + g^*(t)b,\\
\nonumber i\dot{b} &=& g(t) a.
\label{arb}
\end{eqnarray}
Let ${\bm C}$ be a contour that goes around the infinite time semicircle connecting $t \rightarrow \pm \infty $ asymptotics at the real axis as we discussed before. 
If for some reason the amplitude $a(t)$ is known along ${\bm C}$ with initial conditions $(a,b)=(1,0)$ at $t \rightarrow -\infty$, then one can also connect asymptotic values for $b(t)$  as
\beq
b(t\rightarrow+\infty) = -i\int_{\bm C} dt' g(t') a(t'). 
\label{second1}
\eeq

Applying this idea to the extremal amplitude of an LZC model, one would find, however, that it is not enough to know the leading terms in $1/t$ for the amplitude $a(t)$ along ${\bm C}$ because subleading terms in the expansion of $a(t)$ over the small parameter $1/t$  generally produce a finite contribution to (\ref{second1}). 
Nevertheless, imagine that one can find a formal solution for the amplitude $a(t)$ of an extremal state in an LZC-type model  in terms of a formal series in powers of the small parameter $1/t$ along ${\bm C}$. One can then directly substitute this formal solution in expressions like (\ref{second1}) and,
after evaluating Gaussian integrals over time, obtain the series that determines the asymptotic values of other elements of the evolution matrix. So the problem reduces to the question whether interesting LZC models can be found for which not only leading asymptotics but the whole series in powers of $1/t$ can be written explicitly, i.e. not in terms of recursion relations but rather  in terms of explicit expressions for coefficients of this expansion over $1/t$, e.g. in the form of the Tailor series for the generalized hypergeometric function.

The Stokes phenomenon in systems of differential equations with polynomial in time coefficients  has been extensively discussed in mathematical literature \cite{math}. However,  mathematical results  have been  usually formalized to include  too general equations that lack a transparent physical interpretation. In contrast, the major goal of exactly solvable models in physics is to obtain the intuition about a complex nonperturbative regime.  Hence, valuable formulas must be written in terms of physically measurable characteristics, such as transition probabilities. Usually most interesting exact results  can be expressed via elementary functions of model's parameters. It can be also useful when a solvable model can produce  an insight into numerically challenging situations with a macroscopic number of interacting states ($N\gg1$). 

We hope that explicit examples that we provided  will raise the interest in quantum mechanical properties of LZC modes, and this article will be used as the bridge between mathematical literature and physically interesting applications.

\begin{acknowledgments}
Author thanks V. L. Pokrovsky and A. Saxena for useful discussions and M. Anatska for encouragement. 
 This work was funded LDRD and by DOE under Contract No.\  DE-AC52-06NA25396.
\end{acknowledgments}

\appendix

\section{Multistate Landau-Zener models with linear level crossings}
Consider the Hamiltonian with linear time dependence of parameters:
  \begin{equation}
\hat{H}=\hat{A}+\hat{B}t.
   \label{mlzm}
   \end{equation}
 
 Since the Hamiltonian (\ref{mlzm}) has no singularity at $t=0$, typically the scattering problem is formulated for the evolution from $t\rar - \infty$ to $t \rar +\infty$, and we will focus on this case here too. 
 
  \subsection{Brundobler-Elser Formula and No-Go Theorem} 
 It was observed, initially in numerical simulations \cite{be}, that  for any model of the form   (\ref{mlzm}) there are elements of the transition probability matrix, for evolution during $t\in (-\infty, +\infty)$,
  that can be found by a simple application of the two state Landau-Zener formula at every intersection of diabatic energies. Ref.~\cite{be} presented
  a formula for the diagonal element of the scattering matrix for the state whose diabatic energy level has the extremal slope, i.e. if
  $k$ is the index of the state with $B_{kk}=\max(\beta_{1} \ldots \beta_{N})$ or $B_{kk}=\min(\beta_{1} \ldots \beta_{N})$ then
  \begin{equation}
  |S_{kk}(+\infty,-\infty)|=\exp \left( - \pi \sum \limits_{i\,(i \ne k)} \frac{|A_{ki}|^2}{|\beta_k-\beta_i|} \right).
  \label{eq2ng}
  \end{equation}

In  \cite{no-go} another exact result, called the ``no-go theorem", was found in the case when instead of one state with the highest (or one lowest) slope of the diabatic energy level there is a band of an arbitrary number of states having the same 
  highest slope  so that diabatic energies in this band are different only by constant energy parameters. The no-go theorem states that the counterintuitive transitions, as they are defined in the main text,  are exactly forbidden:
    \begin{equation}
 P_{n \rar m} \equiv  |S_{mn}(-\infty,+\infty,)|=0.
  \label{eq3ng}
  \end{equation}
One can easily verify that (\ref{eq2ng})-(\ref{eq3ng}) are direct consequences of the rules (\ref{connect4})-(\ref{connect42}) applied to the systems with the Hamiltonian 
(\ref{mlzm}). 
We note also that validity of (\ref{eq2ng})-(\ref{eq3ng}) was proved by an alternative approach in \cite{mlz-1}. 

\subsection{Discrete symmetries in systems with linear level crossings}
Discrete symmetries of evolution equations with the Hamiltonian (\ref{mlzm}) can be useful to derive constraints on transition probabilities. Here we will show two examples. 

First, consider the class of models of transitions on a linear chain \cite{chain} with evolution of amplitudes $a_{n}(t)$, $n=1,\ldots, N$, of the form:
\beq
i\dot{a}_{n} = \beta_n t a_n +g_n a_{n+1} + g_{n-1}a_n, \quad g_0=g_N=0.
\label{chain11}
\eeq
As we discussed in section IV.C, this system is symmetric under the sign change  of time and, simultaneously, the sign change of amplitudes with even indexes. 
Applying this symmetry to off-diagonal elements of the scattering matrix for evolution from $t\rar -\infty$ to $t \rar +\infty$, we find that $s_{ij}=\pm s_{ji}^*$. 
The latter symmetry means that the transition probability matrix is symmetric:
\beq
p_{i\rar j} = p_{j \rar i},
\label{sym1}
\eeq
which explains some observations in \cite{chain}.

Second, consider a 3-state model with equal couplings between any pair of states:

\beq
 \hat{H}(t) =   \left( \ba{ccc}
\beta t & g& g  \\
g& 0 & g  \\
g& g & -\beta t 
\ea \right).
\label{hamm7}
\eeq
Exact solution for this model has not been found. However, this model is symmetric under  three simultaneously applied operations: 

(i) time reversal $t\rar -t$;

(ii) change of indexes, $1 \rar 3$ and $3 \rar 1$;

(iii) complex conjugation of the evolution equation. 

Let the scattering matrix have the form
\beq
\hat{S} (+\infty|-\infty) =   \left( \ba{ccc}
s_{11} & s_{12} & s_{13}  \\
s_{21}& s_{22}  & s_{23}  \\
s_{31}& s_{32} & s_{33} 
\ea \right),
\label{sa1}
\eeq
then applying a  sort of CPT symmetry  (i)-(iii) in combination with unitarity, $S(t_1|t_2)=S^{\dagger}(t_2|t_1)$, we find that 
\beq
\hat{S} (+\infty|-\infty) =   \left( \ba{ccc}
s_{33} & s_{23} & s_{13}  \\
s_{32}& s_{22}  & s_{12}  \\
s_{31}& s_{21} & s_{11} 
\ea \right).
\label{sa2}
\eeq
Comparing (\ref{sa1}) and (\ref{sa2}), we find constraints on transition probabilities: 
\beq
p_{2\rar 1} = p_{3 \rar 2},\quad p_{2\rar 3} = p_{1 \rar 2}. 
\label{cona1}
\eeq
Unfortunately, conditions (\ref{cona1}) and the Brundobler-Elser formula, which provides  two additional constraints, are still insufficient to determine all transition probabilities in this model.

\section{Exactly solvable multi-state LZC-like model with all nonzero pairwise couplings}

Here, we present an exactly solvable system of the type (\ref{mlz}) that admits the possibility of an arbitrary number of interacting states. Its solution contains some of the  results in sections III.C and IV.D as special cases and hence can be considered as independent verification of connection formulas.

 
 Our model has  $\hat{A}=0$, and we assume that elements of the matrix $\hat{C}$ can be factorized as $C_{ij}=q_iq_j$ with $i,j =1,\ldots, N$, where $q_i$ are characteristic coupling constants. Matrix $\hat{B}$ is assumed to be non-degenerate. 
  $\hat{B}={\rm diag} \{\beta_1,\ldots, \beta_N \}$. We will also assume that state indexes are ordered so that $\beta_i>\beta_j$ if $i<j$. 
  Evolution equation for amplitudes $a_n(t)$ of those states can be written in the form:
  \begin{equation}
 i\frac{d}{dt} a_n=\beta_n t a_n+\frac{q_n}{t} u, \quad u = \sum_{m=1}^{N} q_m a_m.
\label{mmod1}
\end{equation}
where $n=1,\ldots N$.  
Matrix $\hat{C}$ has $N-1$ zero eigenvalues and one nonzero eigenvalue 
\beq
E_{+} = \sum_{m=1}^N q_m^2 
\label{eva2}
\eeq
that corresponds to the eigenstate
\beq
|+ \ra  = \frac{1}{\sqrt{E_{+}}} \sum_{m=1}^N q_m |m \ra .
\label{esa2}
\eeq
Our goal will be to find transition probabilities from this special eigenstate of the Hamiltonian at $t \rar 0_+$ to all possible diabatic states.

First, we perform the change of variables: $u\rightarrow t^2 v$ and $\tau =t^2/2$, leading to  
 \begin{equation}
 i\frac{d}{d\tau} a_n=\beta_n  a_n+q_n v, \quad 2 \tau v = \sum_{m=1}^{N} q_m a_m,
\label{mmod2}
\end{equation}
where $n=1,\ldots N$.  
\begin{figure}
\scalebox{0.17}[0.17]{\includegraphics{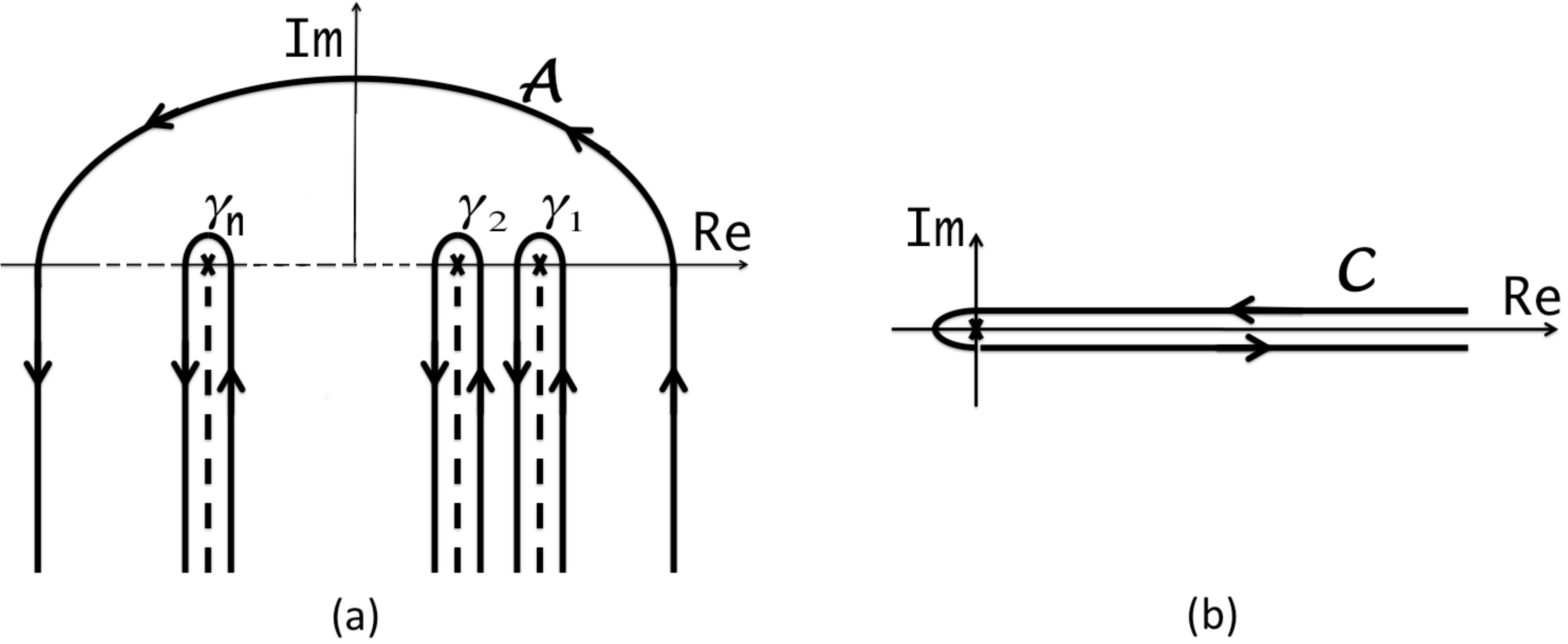}}
\hspace{-2mm}\vspace{-4mm}   
\caption{(a) The integration contour ${\bm A}$ inclosing all branch cuts (dashed lines) from a large distance.  (b) Each integral over ${\bm \gamma_n}$ can be transformed into the integral over the contour ${\bm C}$ by a change of variables.
} \label{cont}
\end{figure}
We introduce the anzatz
\begin{equation}
a_n(\tau)=\int_{{\bm A}} ds e^{-is\tau} \alpha_n(s), \quad v(\tau)=\int_{{\bm A}} ds e^{-is\tau}V (s),
\label{ft1}
\end{equation}
where ${\bm A}$ is a contour such that the integrand vanishes at this contour limits. Substituting (\ref{ft1}) in (\ref{mmod2}) we obtain a 1st order differential equation for $a_n(s)$, which is trivially solvable. Substituting the result back to (\ref{ft1}) we find
\begin{eqnarray}
\nonumber v(\tau) &=& Q\int_{{\bm A}} ds \, e^{-is\tau}  \prod_{n=1}^N (-s+\beta_n)^{\frac{iq_n^2}{2}},\\
 \label{sol11} \\
\nonumber a_{m}(\tau) &=& -Qg_m \int_{{\bm A}} ds \, \frac{e^{-is\tau }}{ -s+\beta_m}  \prod_{n=1}^N \left(-s+\beta_n\right)^{\frac{iq_n^2}{2}},
\end{eqnarray} 
where $Q$ is a normalization constant.

Consider a contour ${\bm A}$, shown in Fig.~\ref{cont}(a), that incloses branch cuts at $s=\{ \beta_n \}$ from a large distance and goes to infinities at $s = -i\infty \pm R$, where $R$ is a large real number. In this limit, we can disregard  terms $\beta_n$ in comparison with $s$, so that integrals in (\ref{sol11}) simplify, e.g., 
\begin{equation}
v(\tau) \rightarrow Q\int_{{\bm A}} e^{-is\tau}  (-s)^{\frac{iE_{+}}{2}} \,d s.\\
\label{sol21}
\end{equation} 
In (\ref{sol21}), the contour ${\bm A}$  can be transformed into the contour ${\bm C}$ in Fig. \ref{cont}(b) by 
switching to the variable $z=iut$ and shrinking the contour to run around the branch cut of $z$. We can then use the formula
\begin{equation}
\Gamma(z) =-\frac{1}{2i{\rm sin}(\pi z)} \int_{\bm C} (-\tau)^{z-1} e^{-\tau} d\tau
\label{GG11}
\end{equation}
to evaluate the integral.
We  find that, at $\tau \rar 0_+$, this solution behaves as the state $|+\ra$, i.e. it corresponds to the desired initial condition if one introduces the normalization factor 
\begin{equation}
|Q|=\frac{1} {\sqrt{4\pi} \sqrt{1-e^{-\pi E_+}}}.
\label{qq1}
\end{equation}
In order to find transition probabilities at $\tau \rightarrow +\infty$ limit, we continuously deform the contour ${\bm A}$ into the combination of contours ${\bm \gamma_n}$ that inclose the branch cuts  at $s=\beta_n$ as shown in Fig. \ref{cont}(a).
In the limit $\tau \rightarrow +\infty$, only the vicinity of the branching points contribute essentially to each integral over ${\bm \gamma_n}$. Hence one can change variables $s \rightarrow u+\beta_n$, keeping the dependence on $s$ only for terms that are singular near the origin of the ${\bm \gamma_n}$-th cut. In all other factors, we can substitute $s$ by its value at this point.  The $m$-th integral in (\ref{sol11}) over  ${\bm \gamma_{m}}$ provides the 
 asymptotic at $t\rightarrow +\infty$ for  $a_{m}(t)$, i.e.
 \begin{widetext}
 \begin{eqnarray}
a_{m}(t)_{\rightarrow+\infty}  =  -Qg_m  \prod_{n=1, \, n \ne j}^N (\beta_n - \beta_m)^{i\frac{q_n^2}{2}}  \int_{\bm \gamma_0} ds \, e^{-ist} (-s)^{-1+i\frac{q_m^2}{2}}, 
\end{eqnarray}  
\end{widetext}
where we should assume that $(-i)=e^{-i\pi/2}$ and $-1=e^{-i\pi}$. Remaining integrals again can be evaluated with Eq.~(\ref{GG11}).
Transition probabilities can be obtained by taking squares of the absolute values of transition amplitudes. In order to write them, it is convenient to introduce LZ-like  probabilities:
\begin{equation}
p_j = e^{-\pi q_j^2}, \quad j=1,\ldots, N.
\label{plz}
\end{equation} 

The transition probabilities from the initially populated  $|+\ra$ state to all possible diabatic states are then given by 
\begin{equation}
\noindent P_{+\rightarrow m}= \frac{\left((1-p_m)  \prod \limits_{n}^{\beta_m<\beta_n} p_n \right)}{1-e^{-\pi E_{+}}}.
\label{p00}
\end{equation}

One can verify that for the special case of $N=2$ and the case when  $m$ is the index of the extremal level, Eq.~(\ref{p00}) transfers into results presented in sections III.C and IV.D. 
We also note that Eq.~(\ref{p00}) predicts that the transition probabilities do not depend on the slopes of the levels $\beta_i$ as far as the ordering of $\beta_i$ according to their magnitudes is preserved.


\begin{thebibliography}{100}


\bibitem{book} H. Nakamura, ``Nonadiabatic Transition", World Scientific Publishing Company, 2nd Edition (2012)



\bibitem{maj} E. Majorana, Nuovo Cimento {\bf 9} (2), 43 (1932)
\bibitem{landau}  L. D. Landau, Physik Z. Sowjetunion {\bf 2}, 46 (1932)
\bibitem{LZ} C. Zener, Proc. R. Soc. A {\bf 137}, 696 (1932); E. C. G. St\"uckelberg, Helv. Phys. Acta {\bf 5}, 369 (1932)

\bibitem{rozen} N. Rosen, and C. Zener,  Phys. Rev. {\bf 40} 502 (1932); J B. Delos and W. R. Thorson, Phys. Rev. A {\bf 6}, 728 (1972); Phys. Rev. A 17, 1343Ð1356 (1978)

\bibitem{nikitin} E. E. Nikitin,  Adv. Quantum Chem. {\bf 5} 135 (1970)


\bibitem{osherov} V. I. Osherov, and H. Nakamura, J. Chem. Phys. {\bf 105}, 2770 (1996)

\bibitem{app-bose}  V. A. Yurovsky, A. Ben-Reuven, and P. S. Julienne, Phys. Rev. A {\bf 65}, 043607 (2002);   V Shahnazaryan, O Kyriienko, I Shelykh, Preprint arXiv/1410.1379 (2014);
W. H. Zurek, U. Dorner, and P. Zoller,
Phys. Rev. Lett. {\bf 95}, 105701 (2005);  B. Damski and W. H. Zurek
Phys. Rev. A {\bf 73}, 063405 (2006);  B. Damski, H. T. Quan, and W. H. Zurek, Phys. Rev. A {\bf 83}, 062104 (2011); V. Gurarie, Phys. Rev. A {\bf 80}, 023626 (2009);  B. E. Dobrescu, and V. L. Pokrovsky, Phys. Letters A {\bf 350},  154 (2006); M. Schecter, and A. Kamenev, Phys. Rev. A {\bf 85}, 043623 (2012)

\bibitem{chain1}  D. Sun and A. Abanov and V. L. Pokrovsky EPL {\bf  83}, 16003 (2008); A. Altland, V. Gurarie, T. Kriecherbauer, and A. Polkovnikov, Phys. Rev. A {\bf 79}, 042703 (2009); A. P. Itin, and P. T\"orm\"a, Phys. Rev. A {\bf 79}, 055602 (2009)


\bibitem{LZ-interferometry} M. N. Kiselev, K. Kikoin and M. B. Kenmoe, EPL {\bf 104}, 57004 (2013); F. Forster {\it et al} Phys. Rev. Lett. {\bf 112}, 116803 (2014);  Sriram Ganeshan, Edwin Barnes, and S. Das Sarma, Phys. Rev. Lett. {\bf 111}, 130405 (2013); Hugo Ribeiro, J. R. Petta, and G. Burkard, Phys. Rev. B {\bf 87}, 235318 (2013)


\bibitem{coher} K. Saito, M. Wubs, S. Kohler, Y. Kayanuma, and P. H\"anggi,
Phys. Rev. B {\bf 75}, 214308 (2007); P. Ao and J. Rammer, Phys. Rev. B {\bf 43}, 5397 (1991) ; M. H. S. Amin, D. V. Averin, and J. A. Nesteroff, Phys. Rev. A {\bf 79}, 022107 (2009); V. N. Ostrovsky and M. V. Volkov,
Phys. Rev. B {\bf 73}, 060405 (2006);  J. Keeling, A. V. Shytov, and L. S. Levitov, Phys. Rev. Lett. {\bf 101}, 196404 (2008);
 M. Wubs, K. Saito, S. Kohler, P. H\"anggi, and Y. Kayanuma, Phys. Rev. Lett. {\bf 97}, 200404 (2006) .

\bibitem{qcontrol} C. M. Quintana, K. D. Petersson, L. W. McFaul, S. J. Srinivasan, A. A. Houck, J. R. Petta, Phys. Rev. Lett. {\bf 110}, 173603 (2013); S. Masuda, K. Nakamura, and A. del Campo, Phys. Rev. Lett. {\bf 113}, 063003 (2014); S. Deffner, C. Jarzynski, and A. del Campo, Phys. Rev. X {\bf 4}, 021013 (2014); A. del Campo, M. M. Rams, and W. H. Zurek, Phys. Rev. Lett. {\bf 109}, 115703 (2012)


\bibitem{be}S. Brundobler and V. Elser, J. Phys. A {\bf 26}, 1211 (1993)

\bibitem{shytov}  A. V. Shytov, Phys. Rev. A {\bf 70}, 052708 (2004)
\bibitem{no-go} N. A. Sinitsyn,  J. Phys. A {\bf 37} (44), 10691 (2004)

\bibitem{mlz-1}    B. E. Dobrescu and N. A. Sinitsyn, J. Phys. B: At. Mol. Opt. Phys. {\bf 39}, 1253 (2006); M. V. Volkov and V. N. Ostrovsky, J. Phys. B: At. Mol. Opt. Phys. {\bf 37}, 4069 (2004); M. V. Volkov and V. N. Ostrovsky,  J. Phys. B: At. Mol. Opt. Phys. {\bf 38}, 907 (2005)


\bibitem{do} Yu. N. Demkov and V. I. Osherov, Zh. Exp. Teor. Fiz. {\bf 53}, 1589 (1967) [Sov. Phys. JETP {\bf 26}, 916 (1968)]; A. A. Rangelov, J. Piilo, and N. V. Vitanov, Phys. Rev. A {\bf 72}, 053404 (2005)

\bibitem{reducible} N. A. Sinitsyn, Phys. Rev. B {\bf 66}, 205303 (2002); J. Dziarmaga, Phys. Rev. Lett. {\bf 95}, 245701 (2005);  M. V. Volkov and V. N. Ostrovsky, Phys. Rev. A {\bf 75}, 022105 (2007)
\bibitem{bow-tie} Y. N. Demkov and V. N. Ostrovsky, Phys. Rev. A {\bf 61}, 032705 (2000); Yu. N. Demkov and V. N. Ostrovsky, J. Phys. B {\bf 28}, 403 (1995); V. N. Ostrovsky and H. Nakamura, J. Phys. A {\bf 30}, 6939 (1997);  Y. N. Demkov and V. N. Ostrovsky, J. Phys. B {\bf 34}, 2419 (2001); C. E. Carroll and F. T. Hioe, J. Phys. A {\bf 19}, 1151 (1986)

\bibitem{chain}  N. A. Sinitsyn,, Phys. Rev. A, {\bf 87}, 032701 (2013); V. L. Pokrovsky and N. A. Sinitsyn, Phys. Rev. B {\bf 65}, 153105 (2002)




\bibitem{coulomb1} V. N. Ostrovsky, Phys. Rev. A {\bf 68}, 012710 (2003)
\bibitem{sinitsyn-13prl} N. A. Sinitsyn, Phys. Rev. Lett. {\bf 110}, 150603 (2013)
\bibitem{sinitsyn-13jpa} J. Lin, and N. A. Sinitsyn, J. Phys. A: Math. Theor. {\bf 47} 015301  (2014)
\bibitem{sinitsyn-14jpa} J. Lin, and N. A. Sinitsyn,  J. Phys. A: Math. Theor. {\bf 47} 175301  (2014)
 
 
 
\bibitem{stark} J. S. Cabral {\it et. al}, New J. Phys. {\bf 12}, 093023 (2010); F. Baumgartner, and H. Helm, Phys. Rev. Lett. {\bf 104}, 103002 (2010); J. M. Menendez, I. Martin, and A. M. Velasco, J. Chem. Phys. {\bf 119}, 12926 (2003);
V. A. Nascimento, L. L. Caliri, A. Schwettmann, J. P. Shaffer, and L. G. Marcassa, Phys. Rev. Lett. {\bf 102}, 213201 (2009); F. Robicheaux, C. Wesdorp, and L. D. Noordam, Phys. Rev. A {\bf 62}, 043404 (2000); Yong-Lin He, J. Phys. B: At. Mol. Opt. Phys. {\bf 45} 015001 (2012)

\bibitem{singular} R. S. Tantawi, A. S. Sabbah, J. H. Macek, and S. Yu. Ovchinnikov  Phys. Rev. A {\bf 62}, 042710 (2000); J. S. Cohen, L. A. Collins, and N. F. Lane, Phys. Rev. A {\bf 17}, 1343 (1978); T. R. Dinterman and J. B. Delos, Phys. Rev. A {\bf 15}, 463Ð474 (1977) 





%
\bibitem{math} V. P. Gurarii and V. I. Matsaev, Teoret. Mat. Fiz. 100 (1994), no. 2,
173Ð182; English transl., Theoret. and Math. Phys. 100 (1994), no. 2, 928Ð936 (1995)




\end{thebibliography}
\end{document}